\documentclass[useAMS,usenatbib,aas_macros]{mn2e}

\usepackage{amssymb,amsmath}
\usepackage{graphicx}
\usepackage{color}
\usepackage{flafter}
\usepackage{placeins}
\usepackage{subfigure}
\usepackage{booktabs}
\usepackage{array}
\usepackage{natbib}
\usepackage{aas_macros}
\usepackage{multirow}
\usepackage[T1]{fontenc} 
\usepackage{aecompl}
\usepackage{csvsimple}
\usepackage{adjustbox}
\usepackage{caption}
\usepackage{hyperref} 

\title[Orbits \& phase-space substructure of an N-body bar]{On the ridges, undulations \& streams in Gaia DR2:\\Linking the topography of phase-space to the orbital structure of an N-body bar}

\author[Fragkoudi et al.]{F. Fragkoudi$^{1}$\thanks{E-mail:
ffrag@mpa-garching.mpg.de}, 
D. Katz$^{2}$, W. Trick$^{1}$, S. D. M. White$^{1}$, P. Di Matteo$^{2}$, 
\newauthor
M. C. Sormani$^{3}$,
S. Khoperskov$^{4,5,2}$,
M. Haywood$^{2}$,
A. Hall\'{e}$^{6,7}$,
A. G\'{o}mez$^{2}$
\\
$^{1}$Max-Planck-Institut f\"{u}r Astrophysik, Karl-Schwarzschild-Str. 1, 85741 Garching, Germany \\
$^{2}$GEPI, Observatoire de Paris, PSL Research University, CNRS, Place Jules Janssen, 92195,
Meudon, France\\
$^{3}$Universit\"{a}t Heidelberg, Zentrum f\"{u}r Astronomie, Institut f\"{u}r theoretische Astrophysik, Albert-Ueberle-Str. 2, 69120 Heidelberg, Germany\\
$^{4}$Max Planck Institute for Extraterrestrial Physics, 85741 Garching, Germany\\
$^{5}$Institute of Astronomy, Russian Academy of Sciences (INASAN), Pyatnitskaya st., 48, 119017 Moscow, Russia\\
$^{6}$Observatoire de Paris, LERMA, CNRS, PSL Univ., UPMC, Sorbonne Univ., F-75014, Paris, France\\
$^{7}$Collège de France, 11 Place Marcelin Berthelot, 75005 Paris, France\\
}

\begin{document}

\date{}

\pagerange{\pageref{firstpage}--\pageref{lastpage}} \pubyear{2014}

\maketitle

\label{firstpage}

\begin{abstract}
We explore the origin of phase-space substructures revealed by the second Gaia data release in the disc of the Milky Way, such as the ridges in the $V_{\phi}$-$r$ plane, the undulations in the $V_{\phi}$-$r$-$V_r$ space and the streams in the $V_{\phi}$-$V_r$ plane. We use a collisionless N-body simulation with co-spatial thin and thick discs, along with orbit integration, to study the orbital structure close to the Outer Lindblad Resonance (OLR) of the bar. We find that a prominent, long-lived ridge is formed in the $V_{\phi}$-$r$ plane due to the OLR which translates to streams in the $V_{\phi}$-$V_r$ plane and examine which closed periodic and trapped librating orbits are responsible for these features. We find that orbits which carry out small librations around the $x_1(1)$ family are preferentially found at negative $V_r$, giving rise to a `horn'-like feature, while orbits with larger libration amplitudes, trapped around the $x_1(2)$ and $x_1(1)$ families, constitute the positive $V_r$ substructure, i.e. the Hercules-like feature. This changing libration amplitude of orbits will translate to a changing ratio of thin/thick disc stars, which could have implications on the metallicity distribution in this plane. We find that a scenario in which the Sun is placed close to the OLR gives rise to a strong asymmetry in $V_r$ in the $V_{\phi}$-$V_r$ plane (i.e. Hercules vs. `the horn') and subsequently to undulations in the $V_{\phi}$-$r$-$V_r$ space. We also explore a scenario in which the Sun is placed closer to the bar corotation and find that the bar perturbation \emph{alone} cannot give rise to the these features.

\end{abstract}



\section{Introduction}
\begin{figure*}
\centering
\includegraphics[width=0.98\textwidth]{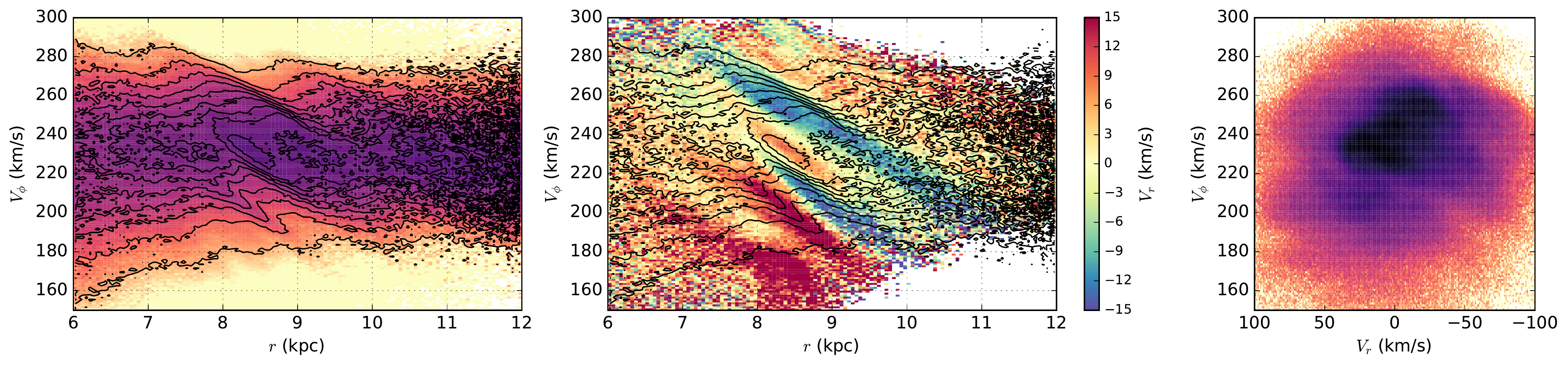}
\caption{Gaia DR2 stars in different planes; we impose quality cuts on parallaxes such that $\frac{\pi}{\sigma_{\pi}} > 5$ and select stars in the disc which have $|z|<500$\,pc. \emph{Left:} Normalised $V_{\phi}$-$r$ plane where the colour-coding indicates the logarithm of density; the plot is normalised according to the total number of stars in each radial bin as in Kawata et al. (2018) to mitigate the effects of the Gaia selection function. \emph{Middle:} The $V_{\phi}$-$r$ plane with the mean $V_r$ in each bin colour-coded -- we can clearly see the undulations in $V_r$ in this plane. The contours are the same as in the plot on the left and show the location of the ridges. \emph{Right:} The $V_{\phi}$-$V_r$ plane for an extended SN of 300\,pc in radius.} 
\label{fig:gaiadr2panels}
\end{figure*}

Ever increasing samples of stars for which we have accurate measurements of positions and velocities have allowed us to continually improve our understanding of the structure of the Milky Way. Well before the release of the Hipparcos catalogue (\citealt{Perrymanetal1997,ESA1997}) which revolutionised our understanding of the velocity distribution of stars close to the Sun and lead to the definitive identification of a number of moving groups, it was hypothesised that clumps in the velocity distribution of stars close to the Solar Neighbourhood (SN) could have a dynamical origin, i.e. be due to resonances induced by non-axisymmetries in the Galactic disc (e.g. \citealt{Kalnajs1991,Weinberg1994}; for a review see \citealt{Antojaetal2010}). 

Often referred to as streams, such as the Hercules stream, these moving groups have been extensively studied in the context of non-axisymmetries caused by the bar and spiral arms (\citealt{Raboudetal1998,Dehnen1999,QuillenMinchev2005}). In the pioneering work of \cite{Dehnen2000}, he showed that one could explain the Hercules stream in terms of the closed periodic orbits found close to the Outer Lindbland Resonance (OLR), if the Sun is placed just outside the OLR, such that $R_{\rm OLR}/R_{\rm 0}\sim0.9$. In this scenario, if the Solar Neighbourhood (SN) is placed at $\sim$8\,kpc, the bar OLR and corotation would be at $\sim7.2$\,kpc and $\sim4$\,kpc respectively (\citealt{Dehnen2000}). This scenario, called the short/fast bar scenario, was also shown to be consistent with  the gas dynamics of the inner Milky Way (e.g. \citealt{EnglmaierGerhard1997} and \citealt{Fux1999}), the value of the Oort constant C and the low velocity moving groups (e.g. \citealt{OllingDehnen2003,Minchevetal2007,Minchevetal2010}).

More recently, studies of star counts of the inner Milky Way have suggested that the length of the bar might be longer than previously thought (\citealt{WeggGerhard2013}) and have lent support to the so called long/slow bar scenario. In this scenario, the thin bar of the Milky Way may extend up to 5\,kpc (\citealt{WeggGerhard2013}; \citealt{Portailetal2017a}), which would make it longer than the corotation radius of $\sim$4\,kpc in the classical short/fast bar scenario. Theoretical studies however show that bar-supporting $x_1$ orbits only extend up to corotation, and as such the bar cannot be longer than its corotation radius (\citealt{Contopoulos1980}), while \cite{Athanassoula1992b} found that the ratio of bar corotation radius to bar length is of the order of $\mathrm{R_{\rm CR}}/\mathrm{R_{\rm bar}} = \mathcal{R}=1.2$. In this scenario of a longer Milky Way bar, the corotation would be found at $\sim$6\,kpc, which would also place the bar OLR further out, at about 10.5\,kpc \citep{Portailetal2017a}. In this case the OLR would be beyond the SN, and the features seen in the velocity distribution close to the Sun could not be explained as being due to orbits at the OLR of the bar. \cite{PerezVillegasetal2017} showed that in such a scenario, orbits affected by the bar corotation could reach the SN and produce a slightly skewed velocity distribution, with a Hercules-like feature, albeit much less pronounced than the feature produced by the bar OLR (e.g. see \citealt{HuntBovy2018,Monarietal2018}). Recent gas dynamical models have also lent support to a long/slow bar model showing that features in the $l-V$ plane can be explained in such a scenario (e.g. \citealt{Sormanietal2015b,Lietal2016}) while recent studies using the continuity equation (i.e. an adaptation of the Tremaine-Weinberg method; \citealt{TremaineWeinberg1984}) find values of the bar pattern speed clustering around $\sim$40\,km/s/kpc (e.g. \citealt{Sandersetal2019, Clarkeetal2019, Bovyetal2019}).
As such, considerable debate still remains as to which of these two scenarios is the correct one for the bar of the Milky Way. 

The second Gaia data release (Gaia DR2; \citealt{Gaia2016,GaiaCollaboration2018}) provided the full 6-dimensional phase-space information of 7.2 million stars brighter than $G_{RVS} = 12$ mag, revealing a plethora of substructures in the disc of the Milky Way (\citealt{GaiaCollaborationKatz2018}). Apart from bringing into focus and revealing more details in the known substructures and moving groups in the Solar Neighbourhood (see e.g. \citealt{Ramosetal2018} and right panel of Figure \ref{fig:gaiadr2panels}), which were well established after the release of the Hipparcos catalogue (e.g. \citealt{Perrymanetal1997,Asiainetal1999,DehnenBinney1998}), new features were also discovered; large-scale ridges were uncovered in the $V_{\phi}$-$r$ plane (\citealt{Kawataetal2018,Antojaetal2018} and see the left panel of Figure \ref{fig:gaiadr2panels}; see also \citealt{Monarietal2017}) which have been associated to wiggles in the rotation curves of galaxies \citep{MartinezMedinaetal2018}, as well as to internal disc instabilities, such as spirals (e.g. \citealt{Huntetal2018}), or satellite interactions (e.g. \citealt{Laporteetal2019}). Superimposed on this plane, we see that there are undulations in $V_r$, i.e. the radial velocity changes sign from positive to negative and back again (see the middle panel of Figure \ref{fig:gaiadr2panels}), which were also observed by \cite{Ramosetal2018} in the moving groups. The observed asymmetry in $V_r$ can also be clearly seen in the Gaia DR2 action space as shown in \cite{Tricketal2018}, and as we see in Figure \ref{fig:gaiadr2panels}, these variations in $V_r$ are correlated to the ridges in the $V_{\phi}$-$r$ plane. Additionally, a number of arches were also identified in the $V_{\phi}-V_r$ plane (\citealt{Antojaetal2018, Ramosetal2018}) which had previously been predicted and associated to a phase-wrapping origin (e.g. \citealt{Minchevetal2009,Gomezetal2012}).

In this paper we explore the phase-space substructures in the $V_{\phi}$-$V_r$-$r$ space using an N-body simulation, where the distribution function and potential are obtained from the self-consistent evolution from axisymmetric initial conditions. This simulation was previously used to study the bulge and inner disc of the Milky Way and was shown to provide a very good fit to these regions (\citealt{Fragkoudietal2017c,Fragkoudietal2018}). We find that a number of features similar to those seen in the Milky Way with Gaia DR2 can be seen in the model and we explore their origin by examining its orbital structure, using spectral analysis and by calculating the closed periodic and trapped orbits.

\begin{figure*}
\centering
\includegraphics[width=0.89\textwidth]{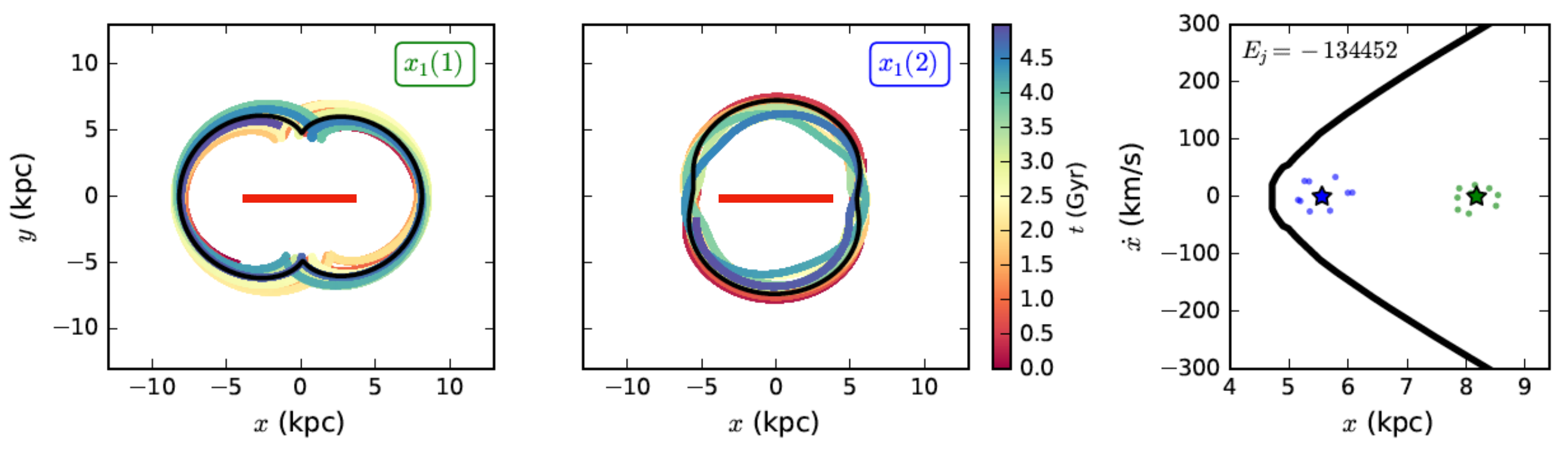}
\caption{\emph{Left:} Closed periodic orbit (solid black line) of the family $x_1(1)$ and trapped orbit (coloured line) of the same energy ($E_J$ = -134452), where the colour coding indicates time. The horizontal red line indicates the bar orientation. \emph{Middle:} As in right panel but for a closed periodic orbit of the family $x_1(2)$ corresponding to the blue star in the surface of section plot in the right panel. \emph{Right:} Surface of section diagram for a given energy $E_J$ = -134452, showing the closed and trapped periodic orbits in the left (green) and middle (blue) panels.} 
\label{fig:cpotrapped}
\end{figure*}

The structure of the paper is as follows: in Section \ref{sec:resorbs} we briefly remind the reader of some important concepts related to closed periodic orbits close to the OLR and trapped orbits around them which are relevant for the rest of the paper. In Section \ref{sec:model} we describe the N-body simulation, the orbit integration method and the frequency analysis of the orbits used in this study. In Section \ref{sec:vphir} we discuss the ridges seen in the $V_{\phi}$-$r$ plane in our model and explore their origin and distribution in energy and angular momentum space. In Section \ref{sec:vphivr} we study the streams in the $V_{\phi}$-$V_r$ plane and link them to the $V_{\phi}$-$r$-$V_r$ space, in terms of the underlying orbital structure, and compare both these planes to Gaia DR2 data. In Section \ref{sec:discuss} we discuss some of the implications of our findings: in \ref{sec:discuss1} we discuss how the metallicity distribution in the $V_{\phi}$-$V_r$ plane will be affected due to the varying libration amplitudes of orbits in this plane, while in Section \ref{sec:discuss2} we briefly explore an alternative scenario where the SN is placed closer to the corotation radius; we find that a model where the SN is placed at corotation cannot explain some of the important features seen in the $V_{\phi}$-$r$-$V_r$ space, specifically the strong asymmetry in $V_r$ in the $V_{\phi}$-$V_r$ plane and the undulations in $V_{\phi}$-$r$-$V_r$. These features are a natural outcome of a scenario in which the Sun is placed just outside the bar OLR, or require the presence of some other non-axisymmetric features, such as spiral arms. In Section \ref{sec:summary} we conclude and summarise our results.

\section{Closed Periodic and Trapped Orbits}
\label{sec:resorbs}

In flattened non-axisymmetric systems such as barred galaxies resonances are induced in the system by the non-axisymmetric bar potential which rotates like a solid body at a given angular frequency, $\Omega_{\rm p}$, often referred to as the bar pattern speed.
A common way of exploring the orbital structure of such systems has been by studying closed periodic orbits (CPO's) in a frame of reference co-rotating with the bar (e.g. \citealt{Contopoulos1980,Athanassoula1992a,Dehnen2000,Patsisetal2002,Sormanietal2015,Fragkoudietal2015}). Closed periodic orbits are in resonance with the bar, i.e. they satisfy the commensurability relation,

\begin{equation}
l\kappa +m(\Omega - \Omega_{\rm p}) = 0.
\end{equation}
They therefore close up upon themselves after $m$ radial and $l$ azimuthal excursions in the co-rotating frame and have dynamical frequencies which are commensurable. 

In barred potentials, contrary to the axisymmetric case, the energy and angular momentum are not integrals of motion, i.e. they are not conserved along a given orbit; the Hamiltonian is however conserved and its constant value, often referred to as the Jacobi energy can be written as,

\begin{equation}
E_J = E - \Omega_{\rm p} L ,
\end{equation}

\noindent and can be thought of as the energy in the rotating frame of reference \citep{BT2008}.
Since $\Delta E_J = 0$ along a given orbit, a change in energy will lead to an equivalent change in angular momentum, such that  
$\Delta E = \Omega_{\rm p} \Delta L$.

Closed periodic orbits are generally separated into stable and unstable; stable closed periodic orbits constitute the backbone of the orbital structure of barred galaxies since they can trap around them non-closed orbits, which will librate around them\footnote{The trapped non-closed orbits can be further subdivided into regular and sticky chaotic (see for example \citealt{SparkeSellwood1987}) although as we discuss below we do not distinguish between these two categories.}. In the left and middle panels of Figure \ref{fig:cpotrapped} we see examples of trapped orbits (coloured lines) of a given Jacobi energy, $E_J$, close to the OLR, which librate around their parent CPO (black lines). These are calculated in our barred galaxy N-body model which is described in the following subsection. These orbits are stable and belong to the $x_1(1)$ and $x_1(2)$ families (nomenclature as in \citealt{ContopoulosGrosbol1989}) and are 2:1 orbits, i.e. they carry out two epicyclic excursions in one rotation around the galactic centre. In the right panel we show a surface of section for these orbits, i.e. we plot the phase-space coordinates $x-\dot{x}$ of the orbit every time it crosses the $y=0$ plane with $\dot{y}<0$. 
The periodic orbits are represented by the stars in the Figure, while the trapped orbits are coloured points clustering around these points; the black line corresponds to the zero velocity curve of the given energy, and delineates the region which orbits can to populate. We therefore see that phase-space in such potentials is made up in part of islands of stability with trapped orbits clustering around their stable parent CPO. 

The further away in phase-space the starting point of the trapped orbit is from its parent CPO, the larger its amplitude of libration, and a CPO will generally trap around it orbits of various librating amplitudes. If a closed periodic orbit is unstable, then orbits will not cluster around it in phase-space (but manifolds of chaotic orbits can emanate from them, see e.g. \citealt{RomeroGomezetal2006}).  As the energy and angular momentum of a closed periodic orbit will fluctuate over time, a trapped orbit's energy and angular momentum will also librate around the values of energy and angular momentum of its parent closed periodic orbit. 

\section{Model \& orbital analysis}
\label{sec:model} 

\subsection{N-body simulation}
\label{sec:nbodymod}
\begin{table}
\centering
\begin{tabular}{ l r | c | c | c | c } 
& & $r_D$ (kpc) & $h_z$ (kpc)  & $M$ ($M_{\odot}$) & $n_p$  \\ \hline
&\emph{Cold} & 4.8 & 0.15 & 4.21 $\times$ $10^{10}$ & $5\times10^6$ \\ \hline
&\emph{Interm.} & 2 & 0.3 & 2.57 $\times$ $10^{10}$ & $3\times10^6$  \\ \hline
&\emph{Hot} & 2 & 0.6 & 1.86 $\times$ $10^{10}$ & $2\times10^6$  \\ \hline
&\emph{DM} & 21 & - & 3.7$\times$ $10^{11}$ & $5\times10^6$  \\ \hline
\end{tabular}
\vspace{0.1cm}
\caption{Properties of the unscaled simulation used in this study. From left to right: the characteristic radius, the characteristic height, the total mass and the number of particles in each component.}
\label{tab:info}
\end{table}
The model we explore here is a purely collisionless N-body simulation, first described in \cite{Fragkoudietal2017c}. Here we give only a summary of the properties of the simulation -- the reader can either refer to \cite{Fragkoudietal2017c} or the Appendix \ref{sec:appendix} for more details. The model is a composite disc galaxy with stellar mass and rotation curve compatible to those of the Milky Way (see the middle panel of Figure \ref{fig:rotcurve}), albeit with values of the circular velocity at the Solar Vicinity ($\sim$205\,km/s) lower than the most recent estimates (see for example \citealt{Eilersetal2018,KawataBovyetal2019} who find values closer to $\sim$230\,km/s). 

The vertically continuous stellar populations seen in the Milky Way disc (e.g. \citealt{Bovyetal2012}) are discretised into three co-spatial discs, described by a Miyamoto-Nagai profile \citep{BT2008}, where each disc has a characteristic radius $r_D$. These can roughly be associated -- morphologically, kinematically and chemically -- to the metal-rich thin disc, the young thick disc and the old thick disc seen in the solar vicinity (nomenclature as in \citealt{Haywoodetal2013}). We call these the cold/thin, intermediate and hot/thick disc respectively, to emphasize their different kinematic and morphological properties. The intermediate and hot discs have a combined mass of 50\% of the total stellar mass of the model (in agreement with the mass growth of the MW disc as estimated by \citealt{Snaithetal2014,Snaithetal2015}), and both have shorter scalelengths than the cold disc, thus making these populations more concentrated in the central regions of the disc (as first shown in \citealt{Bensbyetal2011} and then by \citealt{Bovyetal2012,Chengetal2012}). This composite stellar disc is then embedded in a live dark matter halo (modelled as a Plummer sphere) and let to evolve in isolation from an initial axisymmetric configuration in quasi-equilibrium. For a summary of their properties we refer the reader to Table \ref{tab:info} and for further details to the Appendix \ref{sec:appendix} or \cite{Fragkoudietal2017c}. 

After $\sim$1\,Gyr a strong stellar bar forms which transfers angular momentum from the inner regions of the system to the outer disc and dark matter halo (see also \citealt{Fragkoudietal2017b}).
For most of what follows we use one of the snapshots from the end of the simulation, after 6.7\,Gyr of evolution. From this snapshot we extract the particle positions and velocities (which are used in the plots that follow to explore the phase-space of the model in various projections, e.g. $V_{\phi}-r$, $V_{\phi}-V_r$ etc.), the potential and bar pattern speed. We also use these to carry out the orbital integrations described below. The pattern speed is derived by carrying out a spatial and temporal Fourier decomposition of the surface density of the simulation over 2\,Gyr (e.g. as in \citealt{Halleetal2015}) to extract a spectrogram of the simulation. We rescale the snapshot so that the Solar Neighbourhood is at $R_{\rm 0}$ = 8.3\,kpc (e.g. \citealt{Gillessenetal2017}), just outside the bar OLR which is at 7.3\,kpc (i.e. $R_{\rm OLR}/R_{\rm 0} \sim 0.88$). The bar then has a length of approximately 3.5\,kpc (see left panel of Figure \ref{fig:rotcurve}) and a pattern speed of 50\,km/s/kpc (right panel Figure \ref{fig:rotcurve}).

\subsection{Orbit Integration}
\label{sec:orbint}
Due to the sparsity of saved output snapshots for the N-body simulation, we do not use orbits extracted directly from the simulation outputs, but carry out orbit integrations in the frozen potential of the snapshot discussed in the previous Section. We do not smooth the potential, or fit it with an analytic function, as we want to calculate the orbits directly in the N-body potential. This noisy, asymmetric potential with odd modes, has the down-side of making the orbits less regular than in studies using analytic potentials (adding noise to the potential has been shown to increase the irregularity of orbits, e.g. \citealt{SparkeSellwood1987,Habibetal1997}). 
For the purposes of our study the noisy potential is not prohibitive, since we are not interested in orbital stability or chaos. We note however that we do not distinguish between regular and sticky chaotic orbits in what follows. 

From the simulation snapshot we extract the positions and velocities of the star particles which are used as initial conditions for the orbital integration. The gravitational potential of the snapshot is saved on a 3D grid with 200\,pc resolution and the potential is obtained on intra-grid points using a bi-linear interpolation scheme. We integrate the orbits in the reference frame rotating with the bar pattern speed $\Omega_{\rm p}$ = $\Omega_{\rm p} \hat{e}_z$, where the equations of motion are

\begin{equation}
\ddot{x} = -\nabla \Phi - 2(\Omega_{\rm p} \times \dot{r}) - \Omega_{\rm p} \times (\Omega_{\rm p} \times r)
\end{equation}

\noindent where $ -2(\Omega_{\rm p} \times \dot{r})$ and $-\Omega_{\rm p} \times (\Omega_{\rm p} \times r)$ are the Coriolis and centrifugal force respectively \citep{BT2008}. The equations of motion are integrated using a leapfrog kick-drift-kick algorithm with a time-step of $10^{-5}$\,Gyr for 5\,Gyr. In what follows, all plots showing orbits or quantities related to the orbits are derived using this method, i.e. orbit integration in the frozen N-body potential.

\subsection{Calculating Closed Periodic Orbits}

We search for the closed periodic orbits in the potential -- as extracted directly from the N-body snapshot -- using the orbit shooting method described in \cite{SellwoodWilkinson1993} and employed in numerous studies, e.g. (\citealt{Fragkoudietal2015,Sormanietal2015}). 
The orbits we find belong to the families $x_1(1)$ (e.g. middle panel Figure \ref{fig:cpotrapped}) which are aligned with the bar's semi-major axis and $x_1(2)$  (e.g. right panel Figure \ref{fig:cpotrapped}) which are anti-aligned (nomenclature as in \citealt{ContopoulosGrosbol1989}). We cannot however calculate the stability of these orbits due to the very high numerical accuracy required for such an exercise. 
However, thanks to previous works carried out in analytic potentials (e.g. \citealt{Dehnen2000} and \citealt{ContopoulosGrosbol1989}) we know that the stable $x_1(1)$ and $x_1(2)$ orbits do not likely contain loops, and as such in what follows we consider the $x_1(1)$ and $x_1(2)$ orbits without loops to be stable. Additionally, the CPO's we consider in this study are populated in their surface of section diagrams, and as such, are stable orbits.


\subsection{Spectral analysis}
\label{sec:spectral}

\begin{figure}
\centering
\includegraphics[width=0.48\textwidth]{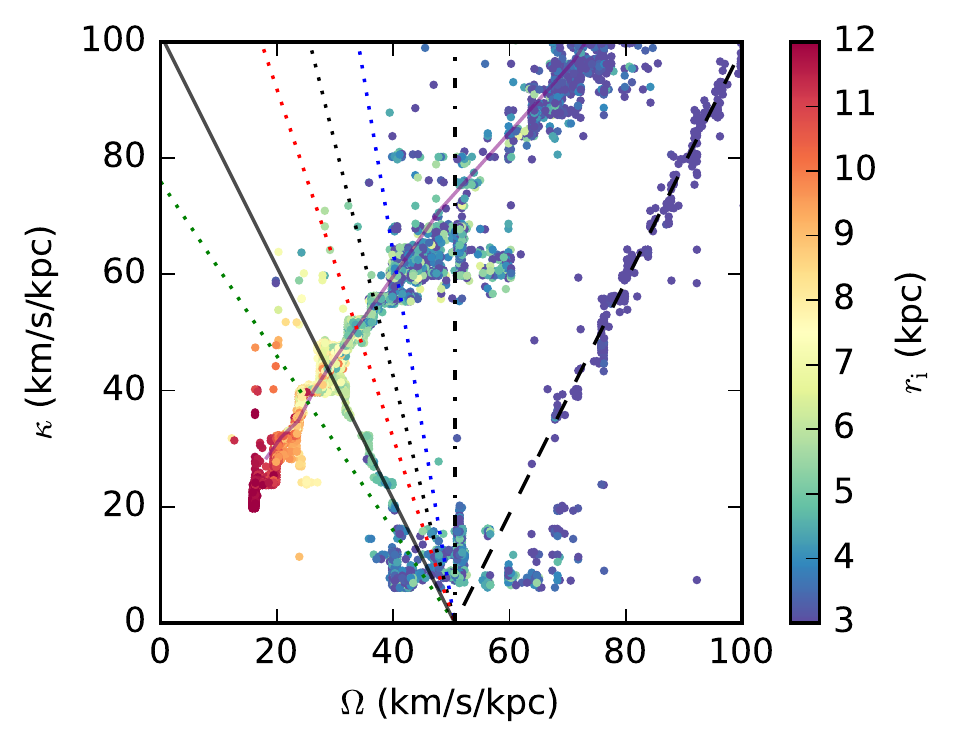}
\caption{Frequency map of 10 000 randomly selected particles in the disc. Some of the main resonances are indicated: the ILR (dashed line), corotation (dashed-dotted line) and the OLR (solid black line). Other higher order resonances are indicated with the dotted coloured lines. The diagonal purple line indicates the relation between $\kappa$ and $\Omega$ obtained from the epicyclic approximation. The colour-coding indicates the inital radius of the particle.} 
\label{fig:freqmap}
\end{figure}

We carry out a spectral analysis of the orbits (e.g. \citealt{BinneySpergel1982,Laskar1993})
in order to obtain the dynamical frequencies $\omega_{\theta}$, $\omega_r$ and $\omega_z$ for each orbit, which describe the frequency of oscillations around the galactic centre, radially and perpendicular to the galactic plane. This is done by recording the position of the orbits at each timestep and computing the fast Fourier transform of the time evolution of cos($\theta$), $r$ and $z$ and extracting the main frequency as the peak with the highest amplitude in the spectrum. For nearly-circular orbits, these correspond to the frequencies obtained with the epicyclic approximation, i.e. $\Omega$, $\kappa$ and $\nu$, and in what follows we use both notations interchangeably. We then use these frequencies to identify orbits which are resonant with the bar, i.e. which have commensurable frequencies with $\Omega_{\rm p}$, as discussed in subsequent sections. 

In Figure \ref{fig:freqmap} we show a frequency map for 10 000 particles randomly sampled from the disc of the model. Overplotted are the lines corresponding to the main resonances, the Inner Lindblad Resonance (ILR) where $\frac{(\Omega - \Omega_{\rm p})}{\kappa} = \frac{1}{2}$ (dashed line), the corotation (CR) resonance where $\Omega - \Omega_{\rm p} = 0$ (dot dashed line) and the OLR where $\frac{(\Omega - \Omega_{\rm p})}{\kappa} = -\frac{1}{2}$ (solid line). Some higher order resonances are indicated with dotted colour lines. The solid diagonal purple line corresponds to the relation between $\kappa$ and $\Omega$ obtained from the epicyclic approximation (\citealt{BT2008}), i.e.,

$$\kappa^2 = \left(r\frac{\mathrm{d}\Omega^2}{\mathrm{d}r} + 4\Omega^2\right).$$

We see that most particles follow this line, while there are also many particles clustered along the resonance lines - the ILR orbits which are the bar-supporting orbits, as well as orbits clustered around corotation and the OLR. The colour-coding in the plot corresponds to the initial radius of the particles.

\section{The $V_{\phi}$ vs $r$ plane}
\label{sec:vphir}
\begin{figure*}
\centering
\includegraphics[width=0.79\textwidth]{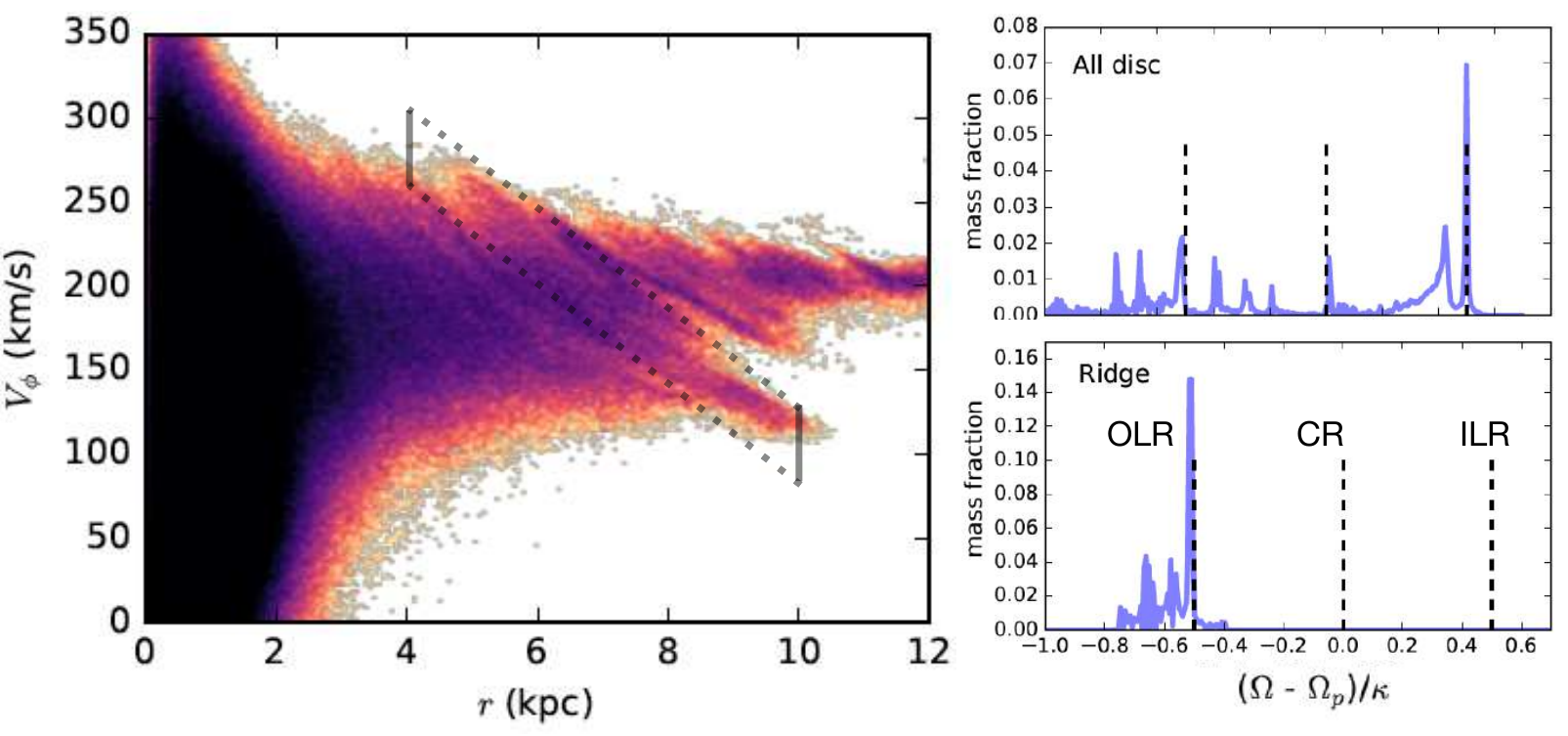}
\caption{\emph{Left:} Logarithmic density map of the $V_{\phi}$ vs $r$ plane in the final snapshot of our N-body model. The dotted box outlines the prominent ridge associated to the bar OLR. \emph{Top Right:} Frequency ratio of orbits for 10 000 particles sampled randomly from the whole disc obtained via orbit integration in the frozen potential from the N-body simulation. \emph{Bottom Right:} Frequency ratio of orbits within the boxed region in the left panel. The vertical dashed lines indicate the three main resonances in barred galaxies, the ILR, the CR and the OLR. We clearly see that the most prominent ridge in the $V_{\phi}$-$r$ plane is made up of OLR orbits. (An animation showing the evolution of the $V_{\phi}$-$r$ plane can be found at \href{https://drive.google.com/file/d/1ySR5LFQGgOd0V3_Y2ZH7_1zS2uHUkm8-/view?usp=sharing}{this link})} 
\label{fig:vphirconsteLz}
\end{figure*}

We now turn our attention to the $V_{\phi}$ vs $r$ plane in our model (see also \citealt{Antojaetal2018,Huntetal2018,Laporteetal2019}). This is plotted in the left panel of Figure \ref{fig:vphirconsteLz}, where the colour-coding shows the density of particles in logarithmic scale. To construct the plot we place the Sun on the $x'$-axis at -8.3\,kpc, i.e. $(x'_{sun},y'_{sun})=(-8.3,0)$ and rotate the bar to be leading the Sun by an angle of 30 degrees (\citealt{Dehnen2000}; and see left panel of Figure \ref{fig:rotcurve}). We then take a slice in $x'$, such that  $|y'|< 2$\,kpc, in order to crudely imitate a selection function similar to that of the Gaia survey. We clearly see that there are prominent ridges in this plane, similar to those seen in the Milky Way in the Gaia DR2 data (e.g. \citealt{Kawataetal2018,Antojaetal2018}) and left panel of Figure \ref{fig:gaiadr2panels}). We remind the reader that as our model is an isolated galaxy, it has not undergone any perturbations from external interactions, and therefore the ridges are the result of the non-axisymmetries in the disc.

There are a number of ridges present in the model, particularly in the outer parts of the disc. By examining their temporal evolution in the N-body simulation\footnote{An animation showing the evolution of the $V_{\phi}$-$r$ plane can be found at \href{https://drive.google.com/file/d/1ySR5LFQGgOd0V3_Y2ZH7_1zS2uHUkm8-/view?usp=sharing}{this link}}, we see that the outermost ridges are undulating and appear transient in nature, and are thus likely associated to weak spiral arms. There is however a prominent ridge centred around $r\sim7$\,kpc, the largest ridge in the plane, which appears shortly after the bar forms and is present at all times thereafter (outlined in the dotted box in Figure \ref{fig:vphirconsteLz}). When it first forms it is located at slightly smaller radii and it moves outwards to its final position at $\sim7$\,kpc.

\subsection{The orbital composition of the $V_{\phi} - r$ plane at the OLR}
\begin{figure*}
\centering
\includegraphics[width=0.98\textwidth]{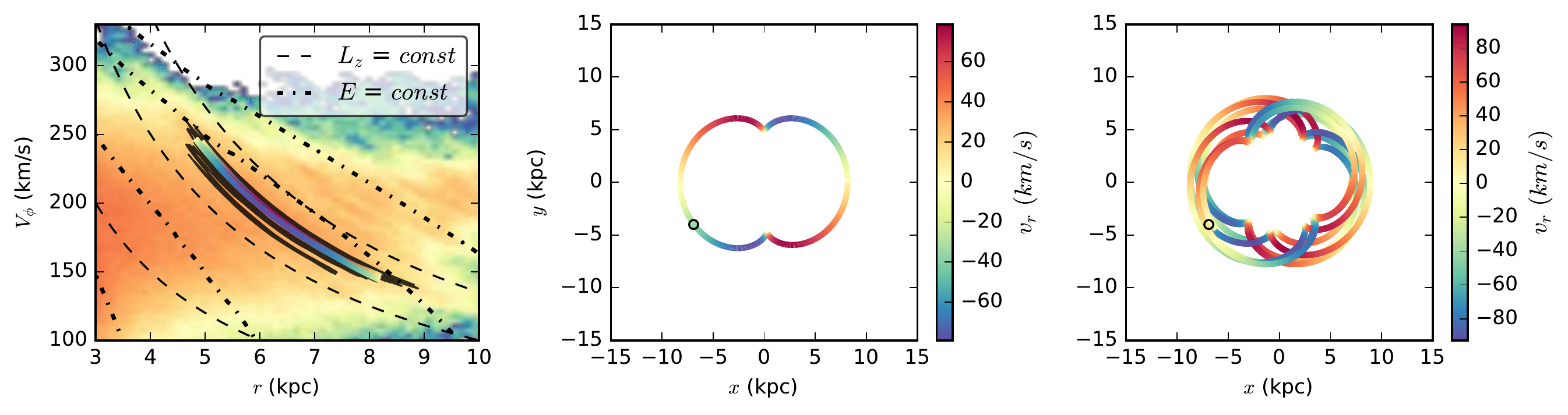}
\includegraphics[width=0.98\textwidth]{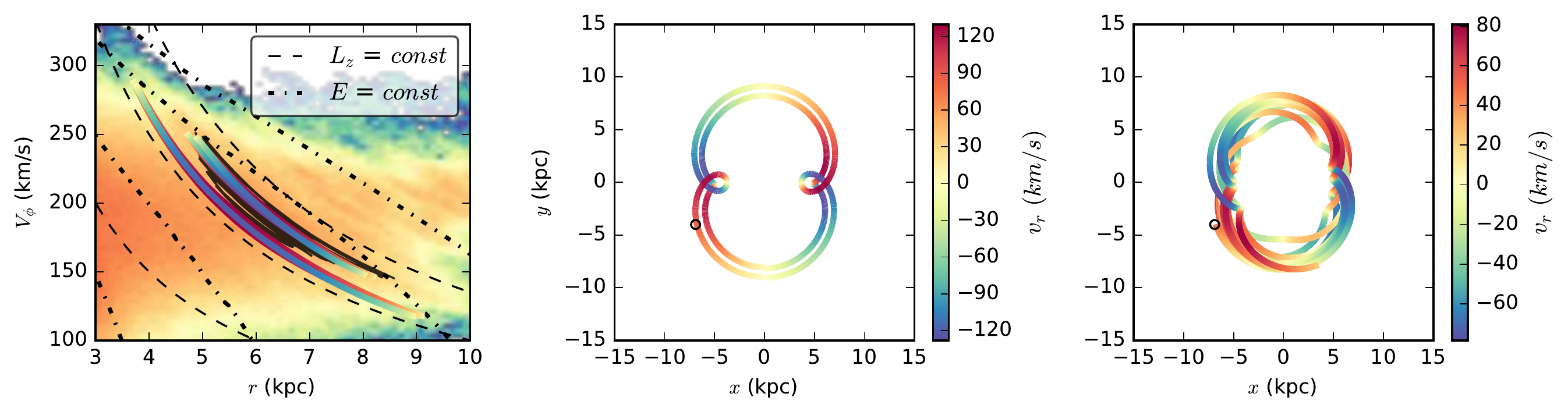}
\caption{\emph{Left:} The $V_{\phi}$ vs $r$ plane from the final snapshot of our N-body model with lines of constant energy and angular momentum overplotted. The colour lines correspond to the path the CPOs (shown in the middle panels) follow in this plane, where the colour-coding corresponds to the $V_r$ velocity. The black lines correspond to the path the trapped orbits (shown in the right panels) follow in this plane. \emph{Middle:} $x$-$y$ morphology of CPO orbits, where the colour corresponds to the radial velocity $V_r$. \emph{Right:} $x$-$y$ morphology of a trapped orbit of the parent CPO shown in the left panel where the colour corresponds to the radial velocity $V_r$. \emph{Top row:} For a CPO of the $x_1(1)$ family. \emph{Bottom row:} For two $x_1(2)$ CPO's, one of which has loops and is likely unstable (i.e. belongs to the $x_1^{\star}(2)$ family). In the middle and right panels the circle denotes the assumed position of the SN and the bar is aligned with the $x$-axis.} 
\label{fig:vphirorb}
\end{figure*}

To investigate the origin of this large and prominent ridge we select all particles inside the dotted box in Figure \ref{fig:vphirconsteLz} and carry out a frequency analysis of their orbits (as described in Section \ref{sec:spectral}). We then compare the orbits in the ridge to the orbits of 10 000 randomly selected particles from the entire disc. This is shown in the right panels of Fig. \ref{fig:vphirconsteLz}; we see that, for the disc particles, there are peaks at frequencies corresponding to the ILR as well peaks at the corotation and OLR resonances (which correspond to $\frac{(\Omega - \Omega_{\rm p})}{\kappa} = \frac{1}{2}, 0, -\frac{1}{2}$ respectively). On the other hand, we see that a large fraction of the particles in the prominent ridge are OLR resonant orbits with $\frac{(\Omega - \Omega_{\rm p})}{\kappa} = -\frac{1}{2}$. We can therefore clearly associate this long and prominent ridge to the OLR of the bar. We note that \cite{Kawataetal2018} identified the longest ridge in Gaia DR2 (at least in terms of its extension to low $V_{\phi}$) as the one associated to the Hercules stream (F1 in their Figure 1).

In the left panels of Figure \ref{fig:vphirorb} we show a zoomed-in region of the $V_{\phi} - r$ plane, with lines of constant energy and angular momentum overplotted. We see that the ridges in this plane appear to follow lines of constant angular momentum (see also \citealt{Laporteetal2019}) at low $V_{\phi}$, and lines of constant energy at high $V_{\phi}$. 
To explore the relation between the closed periodic/trapped orbits and the observed ridges, in the middle panel of Figure \ref{fig:vphirorb} we show closed periodic orbits of the $x_1(1)$ ($x_1(2)$) family in the top (bottom) row and in the right panel we show trapped orbits belonging to these families. In the bottom panel we display two orbits, a stable $x_1(2)$ orbit and an orbit which has loops and is likely unstable (usually referred to as $x_1^{\star}(2)$ orbits; \citealt{ContopoulosGrosbol1989}) to demonstrate how orbits of different radial extents populate the $V_{\phi}$-$r$ plane. The colour coding of these orbits gives the radial velocity $V_r$ along the orbit. 
We overplot these orbits in the $V_{\phi} - r$ plane (left panel of Figure \ref{fig:vphirorb}), with coloured lines for the CPO and black lines for the trapped orbits. We see that the particles move up and down the ridge as they move around their orbit, and that the excursions in angular momentum and energy of the trapped orbits are larger than those of the CPO. Indeed the variations in energy and angular momentum along these orbits is of the order of a few percent, much less than what is seen for example for orbits trapped at corotation (see for example \citealt{Halleetal2018}).

\subsection{The $E-L_z$ plane}

\begin{figure}
\centering
\includegraphics[width=0.48\textwidth]{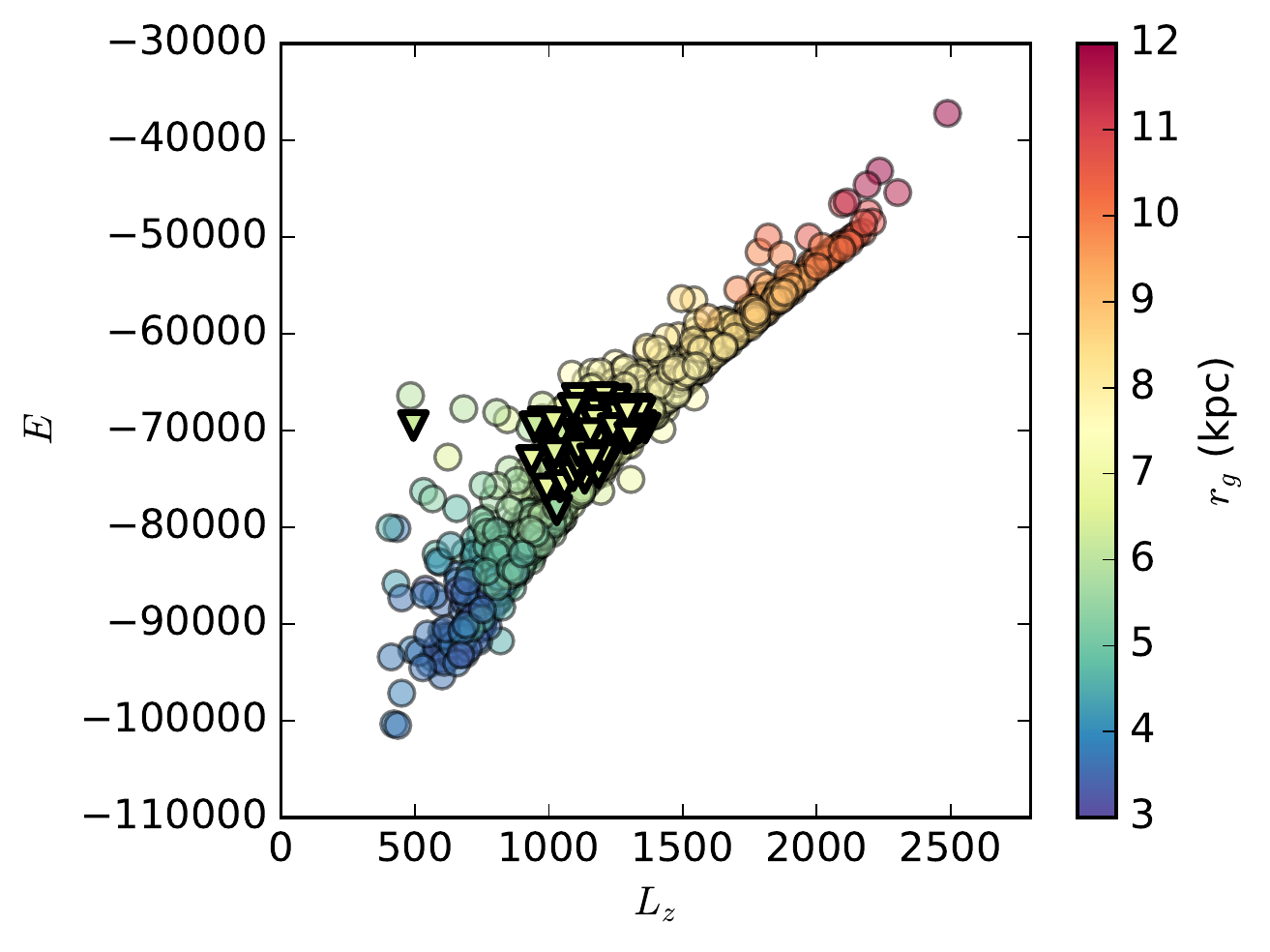}
\caption{Lindblad diagram for particles in the radial range of 4 to 10\,kpc (corresponding to the length of the ridge in Figure \ref{fig:vphirconsteLz}), obtained from the integrated orbits. The circles correspond to all particles while the triangles correspond to particles which are OLR resonant. The colour-coding indicates the guiding radius of the orbits.} 
\label{fig:lindblad}
\end{figure}

In Figure \ref{fig:lindblad} we show a Lindblad diagram ($E-L_z$ plane) for stars with radii between 4 and 10\,kpc which corresponds to the radial extent of the OLR ridge. The circles indicate all stars in this radial range and the triangles are the particles defined as OLR particles, i.e. which have frequency ratio $-0.51<\frac{(\Omega - \Omega_{\rm p})}{\kappa} < -0.49$; these particles have similar guiding radii even though they have instantaneous radii which span a much larger range (i.e. the whole ridge). These stars are clustered in the Lindblad diagram, occupying similar energy and angular momentum regions compared to all particles in that same radial range. This indicates that these particles likely originated with similar energy and angular momenta and that these vary only little due to the non-axisymmetries in the disc; indeed the variation of angular momentum and energy for the OLR particles is on the percent level, much less than what is seen for example for orbits trapped at corotation (e.g. \citealt{CeverinoKlypin2007,MinchevFamaey2010,Halleetal2018}). Since these particles start off being clustered in this space, they remain clustered (modulo the scattering that these particles undergo when the bar first forms and grows and the small variation of angular momentum and energy along their orbit). 

\subsection{Variations of $V_r$ in the $V_{\phi} - r$ plane}

\begin{figure}
\centering
\includegraphics[width=0.49\textwidth]{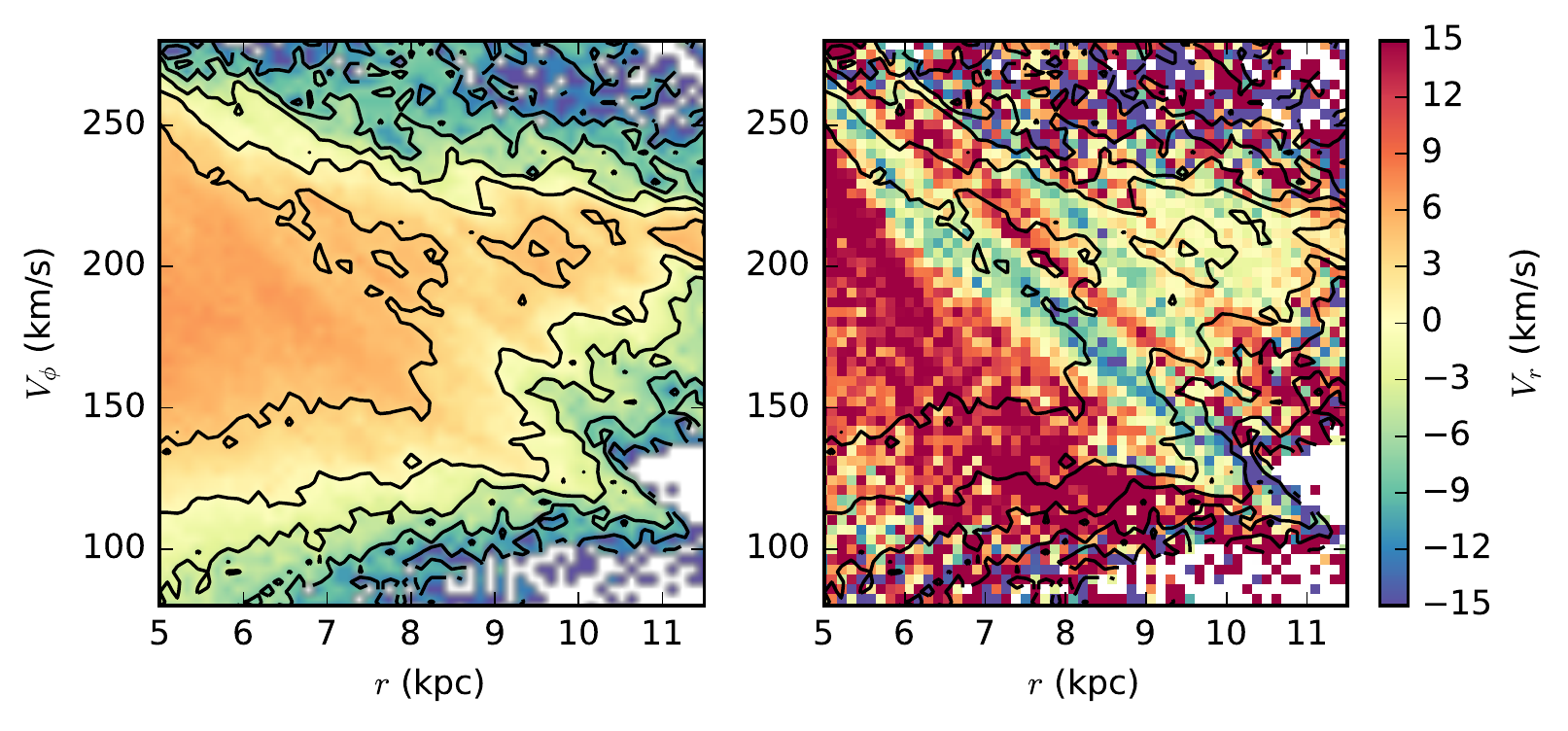}
\caption{\emph{Left:} $V_{\phi}$ vs $r$ density map of the N-body model with contours overploted. \emph{Right:} $V_{\phi}$ vs $r$ plane of the model where the colour-coding shows the mean $V_r$ in each bin. Contours are as in the left plot.} 
\label{fig:vphirandvr}
\end{figure}

We see from the middle panel of Figure \ref{fig:gaiadr2panels} that there are undulations of $V_r$ in the $V_{\phi}$-$r$ plane in the Gaia DR2 data.
While these have been observed in the moving groups (see Figure 6 of \citealt{Ramosetal2018}), here we show these undulations for the first time in the median galactocentric radial motion of $\sim$4.92 million Gaia DR2 stars, selected with $\pi / \sigma_\pi > 5$ and $|z| < 500$~pc (and see \citealt{Tricketal2018} who explored the $V_r$ variations in action space). As can be seen from the Figure, there is a clear connection between $V_r$ and the ridges in the $V_{\phi}$-$r$ plane. Between 8 and 9\,kpc and 180-220\,km/s, following the ridge associated to the Hercules stream (\citealt{Kawataetal2018} and see also \citealt{Monarietal2017}) there is a red outward moving $V_r$ feature. Right next to it and just outside the ridge associated to Hercules, there is an inwards moving $V_r$ feature. There are two further undulations, passing from red to blue which are related to the ridges which have been associated to the Hyades and Pleiades moving groups, and to the Sirius moving group (F3 and F4 as noted in \citealt{Kawataetal2018}).

In our model the longest ridge is related to the OLR of the bar, and we've shown how closed periodic orbits associated to the bar OLR populate this ridge and how $V_r$ varies along their orbit (see Figure \ref{fig:vphirorb}). In Figure \ref{fig:vphirandvr} we show the $V_{\phi}$-$r$  plane on the left and the $V_{\phi}$-$r$ with $V_r$ colour-coded on the right for our model. We see, that as in the Gaia DR2 data, there is a correlation between the ridges in the $V_{\phi}$-$r$ plane and the undulations in $V_r$. In the next Section we explore this relation and we find that the OLR orbits together with the selection function, i.e. the fact that we sample stars at particular points in their orbits, is partly responsible for some of the undulations observed in the $V_{\phi}$-$r$ with $V_r$ space, and the streams in $V_{\phi}$-$V_r$ space at the SN.

\section{The $V_{\phi}$ vs $V_{r}$ plane}
\label{sec:vphivr}

\begin{figure*}
\centering
\includegraphics[width=0.98\textwidth]{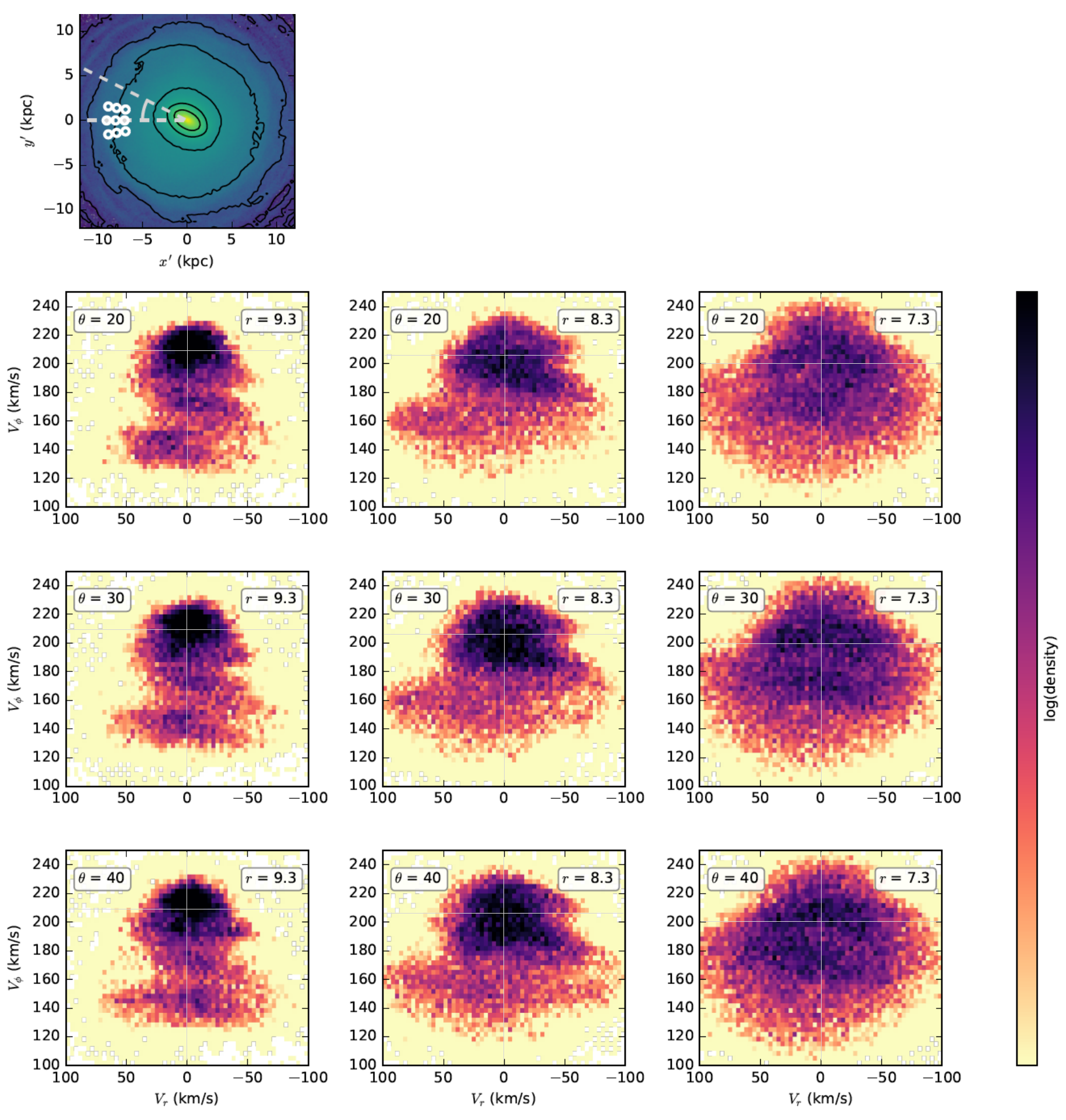}
\caption{$V_{\phi}$ vs $V_r$ for different ``Solar Neighbourhoods'' from stacked snapshots of our N-body model (see text). Top left panel shows the surface density of the model with the locations of the SN's indicated. The angles between the SN's and the bar semi-major axis, as well as the radius of the SN's from the galactic centre are indicated in the top left and right corner of each panel respectively.The fiducial Solar Neighbourhood we study has an angle $\theta=30$ with respect to the bar semi-major axis and distance $r=8.3$\,kpc from the galactic centre. The thin horizontal and vertical lines are added to help guide the eye and correspond to the local value of the circular velocity and the line that passes through $V_r=0$ respectively.} 
\label{fig:vphivrSN}
\end{figure*}

\begin{figure*}
\centering
\includegraphics[width=0.85\textwidth]{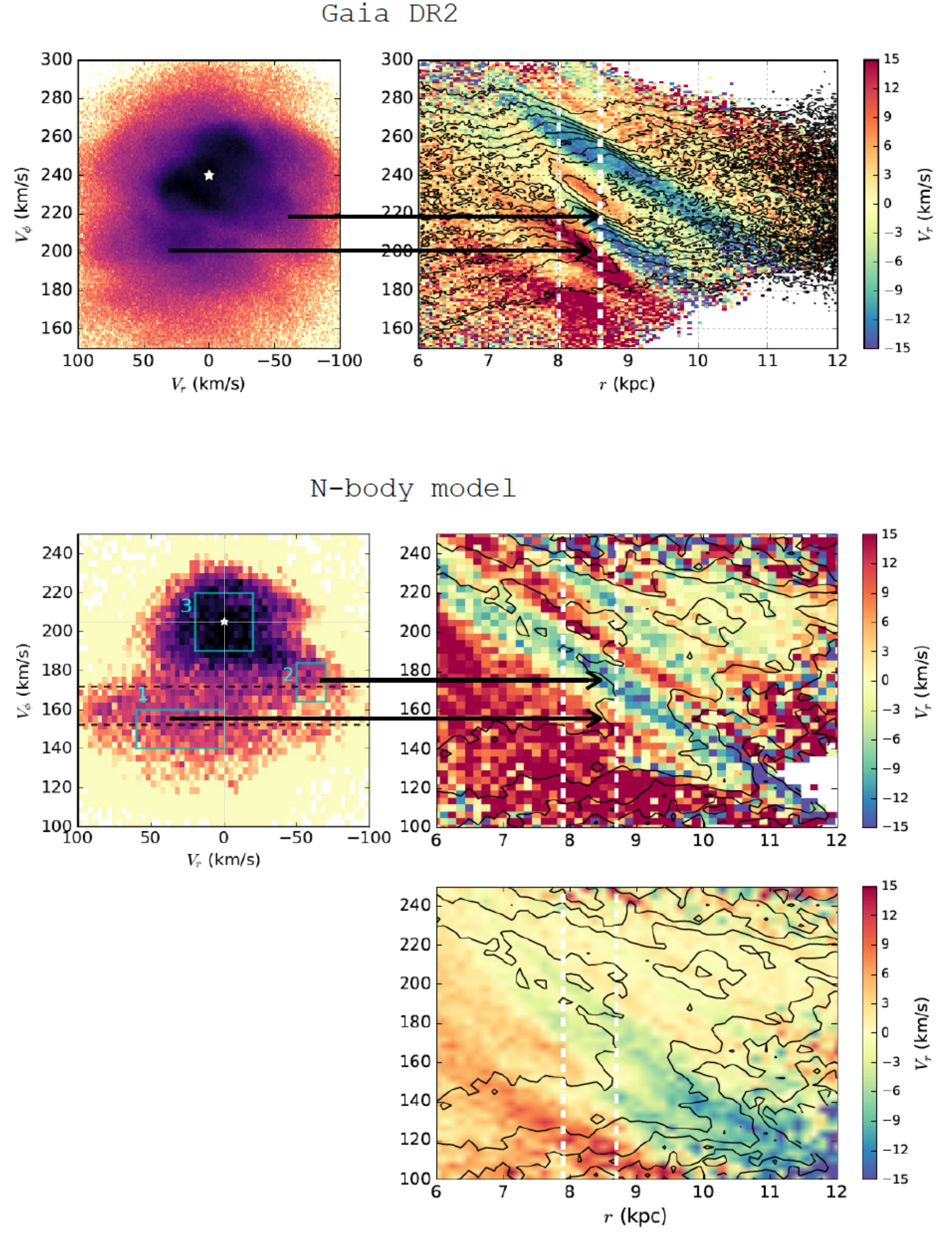}
\caption{\emph{Top panels:} The $V_{\phi}$ vs $V_r$ plane and the $V_{\phi}$ vs $r$ plane with $V_r$ colour coded for the Gaia DR2 data (selection of stars as in Figure \ref{fig:gaiadr2panels}). The arrows point to the location in this plane where there is a switch between inward and outward moving radial velocities which correspond to Hercules and `the horn' in the $V_{\phi}$-$V_r$ plane. \emph{Bottom panels:} The corresponding planes as obtained from our N-body model (see text for details). The arrows point to the two OLR-related structures in our model, i.e. Hercules and `the horn'. The panel in the bottom row shows the $V_{\phi}$ vs $r$ plane with $V_r$ averaged over one full bar rotation, where the bar is kept at the same angle, to wash out any fluctuations in $V_r$ due to e.g. spiral arms; the undulation from outward to inward moving $V_r$ related to Hercules and `the horn' is a robust feature. Note that the $y$-axes of the model and data are not on the same scale due to the lower circular velocity at the Sun in our model as compared to the Milky Way (see Section \ref{sec:nbodymod}).} 
\label{fig:vphiwithvr}
\end{figure*}

We've seen in the previous section that there are correlations between $V_{\phi}$-$r$ and $V_r$. Here we study the $V_{\phi}$-$V_r$ plane at various radii and in detail at our fiducial ``Solar Neighbourhood''.
The stellar streams and substructures evident in the $V_{\phi}$ - $V_r$ plane in the Solar vicinity have been extensively studied and modelled in the literature (e.g. \citealt{Dehnen1998,Dehnen2000,QuillenMinchev2005,Antojaetal2008,Monarietal2016,PerezVillegasetal2017,HuntBovy2018,Monarietal2018}). Here, we revisit this plane in an N-body setup (see also \citealt{Quillenetal2011}), and explore the orbital structure that constitutes its different regions when the Sun is placed just outside the OLR. We also explore this plane in the context of a model with a thin and thick disc, and investigate the variations in the ratio between the density of thin/thick disc stars in it.

In Figure \ref{fig:vphivrSN} we show the $V_{\phi}$-$V_r$ plane for different regions of our model, as indicated in the top left panel of the Figure. For each Solar Neighbourhood we select particles within a radius of 400\,pc and we take all particles belonging to the thin and thick discs. To construct the figure we use 10 snapshots (separated by 10\,Myr each), where we rotate the bar to always have the same position angle with respect to the galactocentric line of site.  We do this to overcome the low signal to noise for a single snapshot, as only a few hundred stars fall in our SN region in a given snapshot. Also, by stacking the different snapshots, this smooths out any short-lived transient phenomena in this plane and will enhance the features due to the bar. Each row shows regions with a given angle with respect to the bar and three different radii, with the angles denoted in the top left corner of each panel and the radius of the region in the top right corner. We see, in accordance with what has been shown in previous studies, that the $V_{\phi}$ - $V_r$ plane morphology is dependent on its location in the disc. This is due to the change in orbital structure in the region just within and outside the OLR. We also see that the change in radius affects the $V_{\phi}$-$V_r$ plane much more than a change in angle -- for variations of the angle of $\sim20$deg -- as was shown to be the case for Gaia DR2 data (e.g. \citealt{GaiaCollaborationKatz2018,Ramosetal2018}). We also see that the stream corresponding to the ``Hercules'' stream moves up in $V_{\phi}$ for smaller radii, and moves down in $V_{\phi}$ for larger radii which is coherent with the diagonal geometry of the OLR ridge associated to it. 

In the top panels of Figure \ref{fig:vphiwithvr} we show the same plots but for an extended Solar Neighbourhood ($d$=300pc) in the Gaia DR2. In the bottom panels we show the $V_{\phi}$ - $V_r$ plane at the SN and the $V_{\phi}$-$r$ plane with $V_r$ colour-coded, of our model. The $V_{\phi}$ - $V_r$ plane is constructed in the same way as in Figure \ref{fig:vphivrSN}, i.e. by stacking 10 snapshots (in this figure we highlight the regions that are studied in detail in the next subsection). We see interesting similarities between the model and the data, in both planes, and we note the striking undulations in the $V_{\phi} -r$ plane with $V_r$ superimposed. Note that the $y$-axes for the model and data do not have the same scale, due to the lower rotational velocity $V_0$ at the Sun in the model.

We first see that the features called the Hercules stream and the `horn'\footnote{The horn has also been identified as the moving groups Wolf 630 and Dehnen98 in \cite{Antojaetal2012}.} in the UV plane are found at the same relative positions in the $V_{\phi}- V_r$ plane of the model as in the data (modulo the difference in circular velocity at the SN in our model), and are prominent long-lived, stable features (these correspond to the regions marked 1 and 2 respectively). In our model, these two features are what are primarily responsible for creating one of the transitions from outward to inward moving radial velocities, $V_r$, in the $V_{\phi} - r$ plane, as indicated by the arrows. Indeed we see in the $V_{\phi}$ - $r$ with $V_r$ plane that this transition from outward (red) to inward moving (blue) occurs close to the `OLR ridge' (as defined in the previous Section). 

By investigating the temporal evolution of the $V_{\phi}$-$r$ with $V_r$ plane we see that there can be a rapid variation in the features, due possibly to the low number of particles in constructing the figure, as well as the inherent noisiness of self-consistent N-body simulations. To isolate the features which are due to the bar, and not caused by spirals or other asymmetries, in the bottom panel of Figure \ref{fig:vphiwithvr} we show the average $V_{r}$ in the $V_{\phi}$-$r$ plane, over a period of one full bar rotation. The figure is constructed by keeping the bar at a fixed angle, and crudely imposing cuts on the spatial selection of stars (of $|y|<2$\,kpc) to mimick the Gaia DR2 selection function, as was done in the previous plots.
This stacked, or smoothed, $V_{\phi}$-$r$-$V_r$ plane allows us to identify which features are due to the bar, and which are short-lived features with different pattern speeds to that of the bar, for example due to spiral arms, since these will be washed out.
We see that a stable long-lived feature in this space, which is a result of the bar-induced non-axisymmetries, is the inward and outward moving part of the OLR ridge. This feature, as we will see in the next Section, is due to the $x_1(1)$ and $x_1(2)$ orbits both existing in this region, which give rise to `the horn' and Hercules streams. Therefore in our model, the bottom undulation seen in $V_r$ in the $V_{\phi} - r$ plane is due to the bar OLR, together with the selection function while the top undulation is likely due to spiral arms.

\subsection{The orbital composition of the $V_{\phi}$ - $V_r$ plane at the OLR}
\begin{figure*}
\centering
\includegraphics[width=0.85\textwidth]{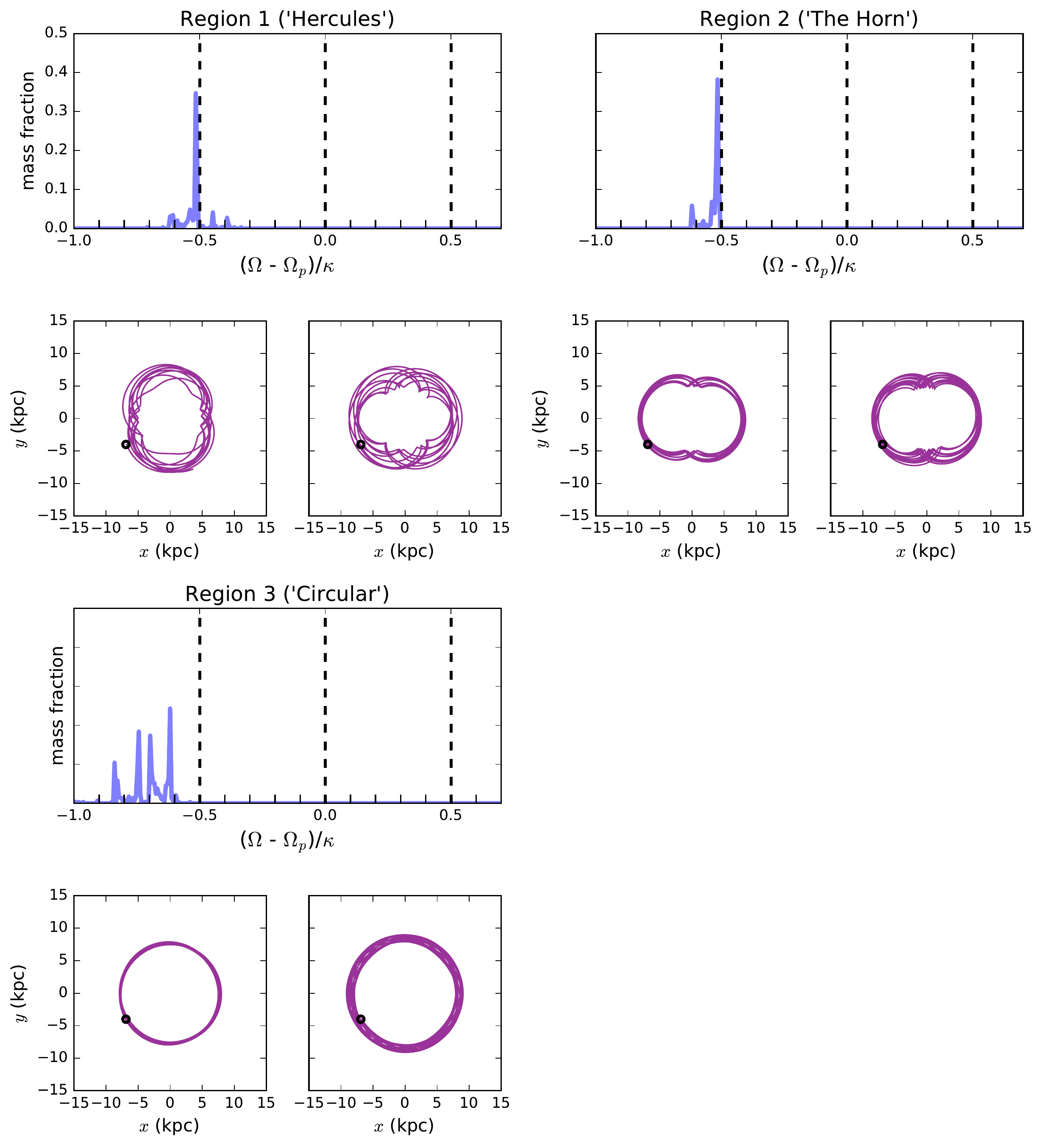}
\caption{Frequency ratios and typical orbits in each of the regions indicated in Figure \ref{fig:vphiwithvr}. The first and third row panels show the frequency ratios for the orbits in each of the regions. The second and fourth row panels show typical orbits in each of the regions. The black circle in these panels indicates the location of the fiducial solar neighbourhood and the bar is oriented along the $x$-axis.} 
\label{fig:regionsfreqorbs}
\end{figure*}

To better understand the origin of the structures seen in the $V_{\phi} - V_r$ plane we explore its orbital composition in some detail. In the bottom left panel of Figure \ref{fig:vphiwithvr}, we highlight three regions: region 1 corresponds to a Hercules-like feature, region 2 corresponds to a `horn'-like feature and region 3 corresponds to almost circular orbits, often referred to as the `LSR mode'. To investigate the origin of these features we select the particles in each of these regions and perform an orbit integration in the frozen potential as described in Section \ref{sec:orbint}. 

The results of this are shown in Figure \ref{fig:regionsfreqorbs}. For each region we show a panel with the orbital frequencies of all the integrated orbits, and below that, two characteristic orbits of each region plotted in the $x-y$ projection where the SN is marked with a black circle. For region 1, which can be associated to the Hercules stream, we see that its orbital frequencies suggest that it is indeed related with the bar OLR in our model, i.e. most orbits satisfy $\frac{(\Omega - \Omega_{\rm p})}{\kappa} = -\frac{1}{2}$. In the second row of Figure \ref{fig:regionsfreqorbs} we show two typical orbits found in region 1. We see that they are orbits librating around the closed periodic orbits that correspond to the $x_1(1)$ (right orbit) and $x_1(2)$ (left orbit) families. Region 3 is essentially made up of orbits on almost circular orbits, as can be seen from the frequency ratios and the typical orbits seen in this region. The orbits in region 2, which corresponds to `the horn',  are also associated to the bar OLR, i.e. there is a sharp peak at $\frac{(\Omega - \Omega_{\rm p})}{\kappa} = -\frac{1}{2}$. 

\begin{figure}
\centering
\includegraphics[width=0.79\linewidth]{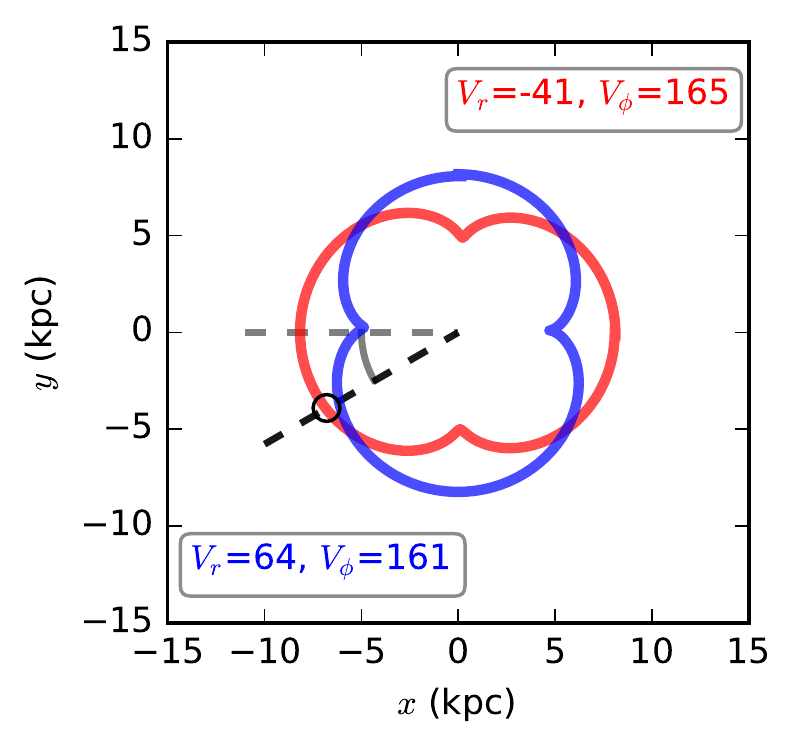}
\caption{2:1 closed periodic orbits which pass through, or close to, the Solar Neighbourhood: The orbits belonging to the  $x_1(1)$ family are shown in red and the $x_1(2)$ family in blue. The inset boxes indicate the $V_{\phi}$ and $V_r$ velocities (red for $x_1(1)$, blue for $x_1(2)$) for the orbits when they cross an angle of 30 degrees (indicated by the dashed black line), i.e. close to the Solar Neighbourhood (indicated by the black circle). The two families contribute to creating the asymmetry in $V_r$.} 
\label{fig:cpoSN}
\end{figure}

\begin{figure*}
\centering
\includegraphics[width=0.99\textwidth]{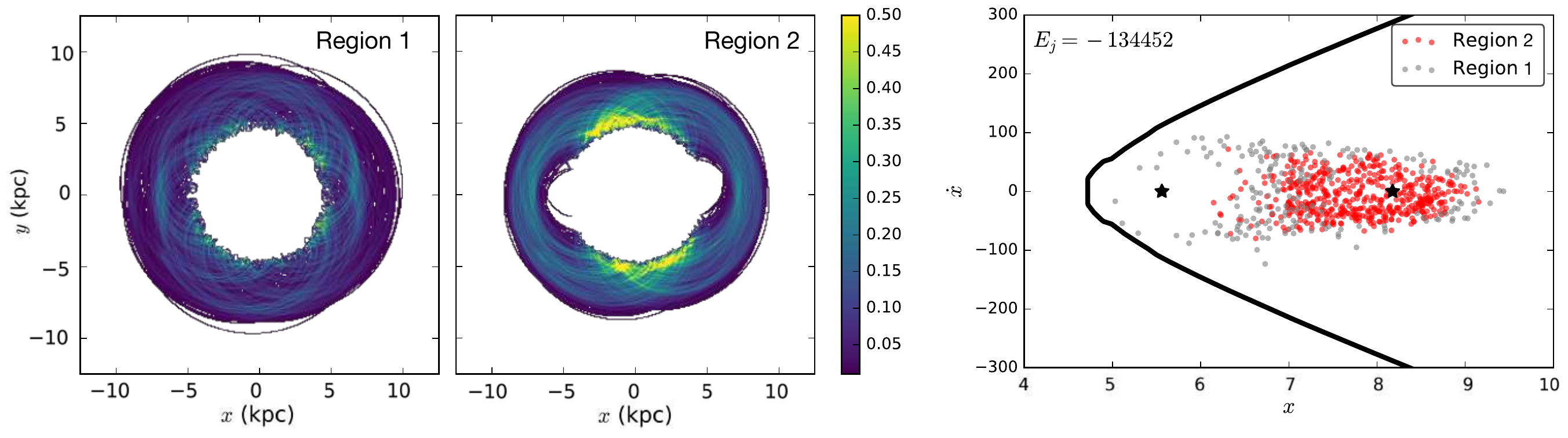}
\caption{\emph{Left and middle:} Surface density of orbits in region 1 and 2 (bar oriented along the $x$-axis) and \emph{right:} a Poincar\'{e} diagram for a specific energy characteristic of OLR orbits. We see that region 2 orbits (red points) are trapped around the $x_1(1)$ parent orbit, and execute small librations. Region 1 orbits (grey points) are trapped around both the $x_1(1)$ and $x_1(2)$ family and have larger libration amplitude than region 2 orbits.} 
\label{fig:SOSregion1region4}
\end{figure*}

How do we obtain two features associated to the OLR but with opposite $V_r$, i.e. Hercules and `the horn'? This can be understood by considering Figure \ref{fig:cpoSN} and Figure \ref{fig:SOSregion1region4}. In Figure \ref{fig:cpoSN} we plot the 2:1 closed periodic orbits which pass through, or close, to the SN in our model; we see that there are both $x_1(1)$ and $x_1(2)$ CPO's that can reach the SN. These CPO's from the two different families, when sampled close to the location of the solar neighbourhood, will contribute to opposite sides of the  $V_{\phi} - V_r$ plane (as noted also in \citealt{Dehnen2000}). The $V_{\phi}$ and $V_r$ velocities of these orbits when they cross the angle between the Sun and the bar semi-major axis, which in our fiducial setup is 30 degrees (as indicated by the black dashed line in the figure) are marked in the top right corner in red for the $x_1(1)$ CPO and in the left bottom corner in blue for the $x_1(2)$ orbit. While the $x_1(2)$ closed periodic orbit does not reach the SN itself, orbits which carry out even small librations around the CPO will reach it. Therefore we see that the $x_1(2)$ CPO and trapped orbits contribute to the outwards moving positive $V_r$, while the $x_1(1)$ CPO contributes to negative inward moving $V_r$ (see also \citealt{Dehnen2000}). 

There is one more point however which we must consider: in Figure \ref{fig:regionsfreqorbs} we see that region 1 is constituted by both $x_1(2)$ and $x_1(1)$ trapped orbits. How do the $x_1(1)$ orbits contribute to the positive $V_r$ region since their CPO contributes to negative $V_r$ when sampled at the SN?
This can be understood by examining Figure \ref{fig:SOSregion1region4}. In the left panel we show the total surface density of stars in region 1, and in the middle panel the total surface density of stars in region 2. We see that particles that make up region 1 are on orbits which have a rounder morphology, while  particles that make up region 2 show a more flattened distribution, akin to the $x_1(1)$ orbital family. This is because region 1 consists of orbits carrying out large librations around \emph{both} the $x_1(1)$ and the $x_1(2)$ families, while region 2 consists predominantly of orbits librating with small excursions around the $x_1(1)$ family.

We confirm this by examining the right panel of Figure \ref{fig:SOSregion1region4}, where we show a surface of section diagram for a given $E_J$, typical of the OLR orbits that make up region 1 and 2. We see that while region 2 particles (red points) only librate around the $x_1(1)$ closed periodic orbit, region 1 particles (grey points) librate around both the $x_1(1)$ and the $x_1(2)$ parent closed periodic orbit. The orbits in region 1 are orbits which are trapped around one family, but we also found cases of orbits which fluctuate between the two families (see also \citealt{Minchevetal2010} who found that orbits precess in setups where the disc undergoes relaxation after introducing the bar perturbation). We also see that the amplitude of libration is smaller for region 2 particles compared to region 1 particles, since they are enveloped within the region 1 particles. The larger librations of $x_1(1)$ trapped orbits in region 1 will result in them having positive $V_r$'s when passing through the solar vicinity. This can also be seen by examining the top middle and right panels of Figure \ref{fig:vphirorb}; we see that while the $V_r$ at the SN for the $x_1(1)$ CPO is negative, the $V_r$ for the highly librating trapped $x_1(1)$ orbit in the right panel can be positive at the SN, as long as the libration around the CPO is large enough, such that the apocentre of the orbit has an angle similar to or larger than the angle between the bar semi-major axis and the galactocentric line of sight.

\section{Discussion}
\label{sec:discuss}

\subsection{Implications on the metallicity distribution}
\label{sec:discuss1}

\begin{figure}
\centering
\includegraphics[width=0.95\linewidth]{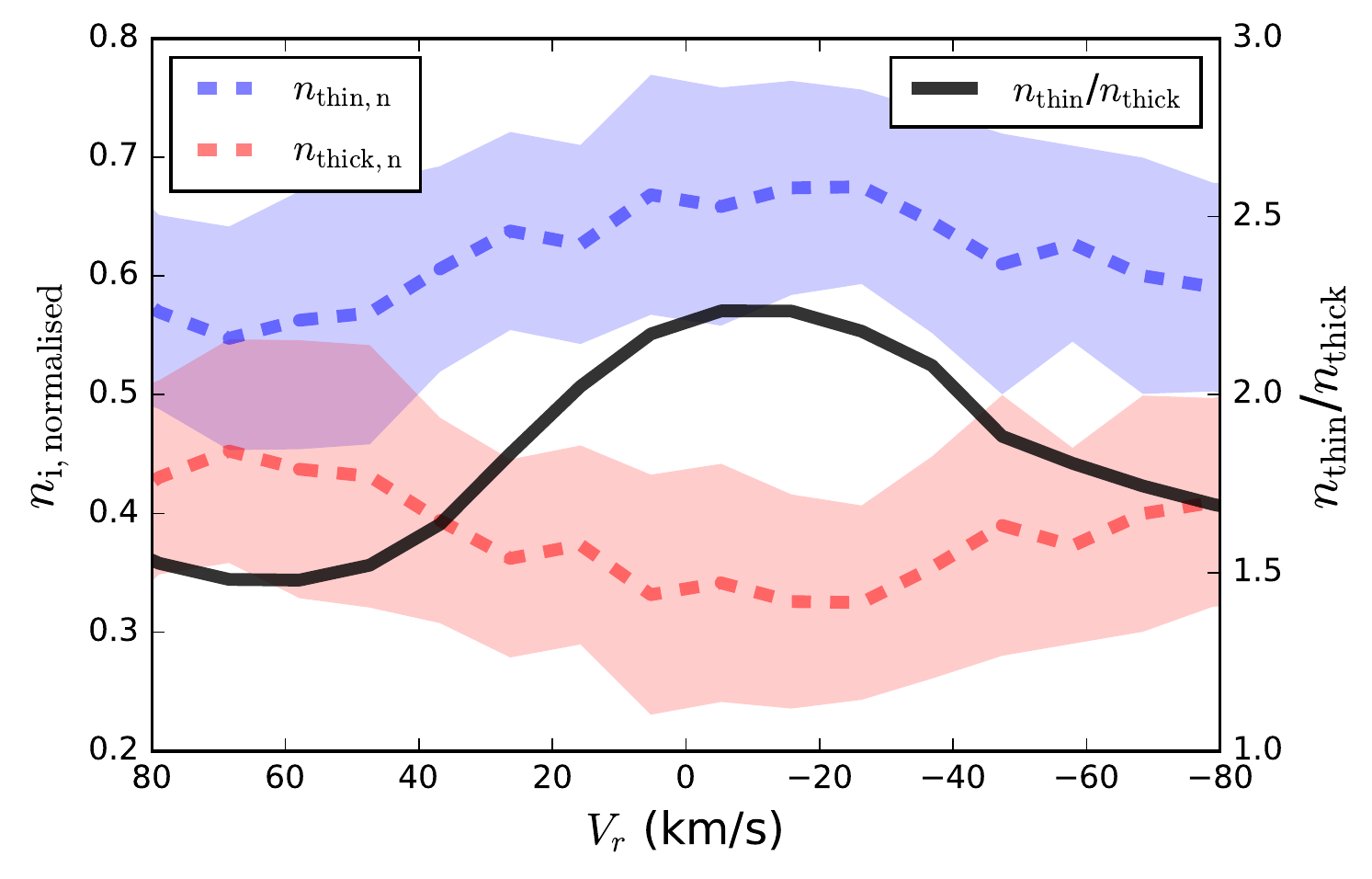}
\caption{Ratio of thin/thick disc particles for $V_{\phi}$ = 162$\pm$10km/s. The dashed blue and red curves show the average number of particles in the thin and thick disc respectively over 1 bar rotation period (the shaded regions shows the 1$\sigma$ dispersion). The solid black line shows the mean value of the ratio of thin to thick disc particles. We see that the region of negative $V_r$ has a higher fraction of thin disc particles as compared to the positive $V_r$ region (which corresponds to the Hercules stream).} 
\label{fig:ratiothinthick}
\end{figure}

In the previous Section we show that orbits in region 1 tend to have larger libration amplitudes around their parent closed periodic orbit than orbits in region 2. By drawing a comparison to the axisymmetric case, the larger libration amplitude suggests that these particles originated from a hotter component, i.e. the particles were originally on orbits with larger epicyclic frequencies. Thus they would likely originate from a hotter component with a large radial velocity dispersion, as compared to region 2 particles. In our model, this population corresponds to the hot/thick disc which is also more metal-poor and $\alpha$-enhanced. Therefore if the different regions of the $V_{\phi}$-$V_r$ plane are populated differently by particles originating in a cold and hot population, this could translate to a noticeable effect in terms of the metallicity distribution in the $V_{\phi} - V_r$ plane. 

For a purely axisymmetric disc setup, the thin/thick disc ratio (and therefore the metallicity) should not be asymmetric in $V_r$ for a given $V_{\phi}$ -- even though the thin/thick disc ratio may change as a function of $V_{\phi}$ itself. However, as hinted at from our exploration of the orbital structure above, the thin/thick disc ratio may well be asymmetric in $V_r$ in a non-axisymmetric setup. As we have both a thin and thick disc in our model, we can explore the thin/thick disc ratio as a function $V_r$ for a given $V_{\phi}$; we show this in Figure \ref{fig:ratiothinthick}. We select particles with $V_{\phi}$ = 162$\pm$10\,km/s (see also the horizontal dashed lines in Figure \ref{fig:vphiwithvr}) in order to select particles which are both in region 1 and region 2. We then calculate the number of thin and thick disc particles, normalised by the total number of particles in each $V_r$ bin, as a function of $V_r$ (over an entire bar rotation to increase the signal to noise). The dashed blue and red lines show the average value of thin and thick disc particles, respectively, normalised by the total number of particles in each $V_r$ bin; the shaded regions show the 1$\sigma$ dispersion. We see that there is a variation of thin/thick disc ratio as a function of $V_r$, which is shown with the solid black line: at negative $V_r$'s corresponding to `the horn' (i.e. region 2) there is an increase in the thin/thick disc ratio as compared to positive $V_r$'s (i.e. the Hercules stream). This is likely due to the fact that particles on colder orbits populate region 2, while particles from the hotter population populate region 1.

Previous work has shown that there are variations in metallicity in the $V_{\phi}$-$V_r$ plane (e.g. \citealt{Antojaetal2017,Hattorietal2018}) and a detailed comparison of our model to the metallicity distribution in this plane will be the subject of a future study. However it is worth noting that `the horn' in the $V_{\phi} - V_r$ plane of the Milky Way, which corresponds to region 2, seems to have higher metallicity than surrounding regions (e.g. \citealt{Antojaetal2017}).

\subsection{The slow vs. fast bar cases}
\label{sec:discuss2}

\begin{figure*}
\centering
\includegraphics[width=0.98\textwidth]{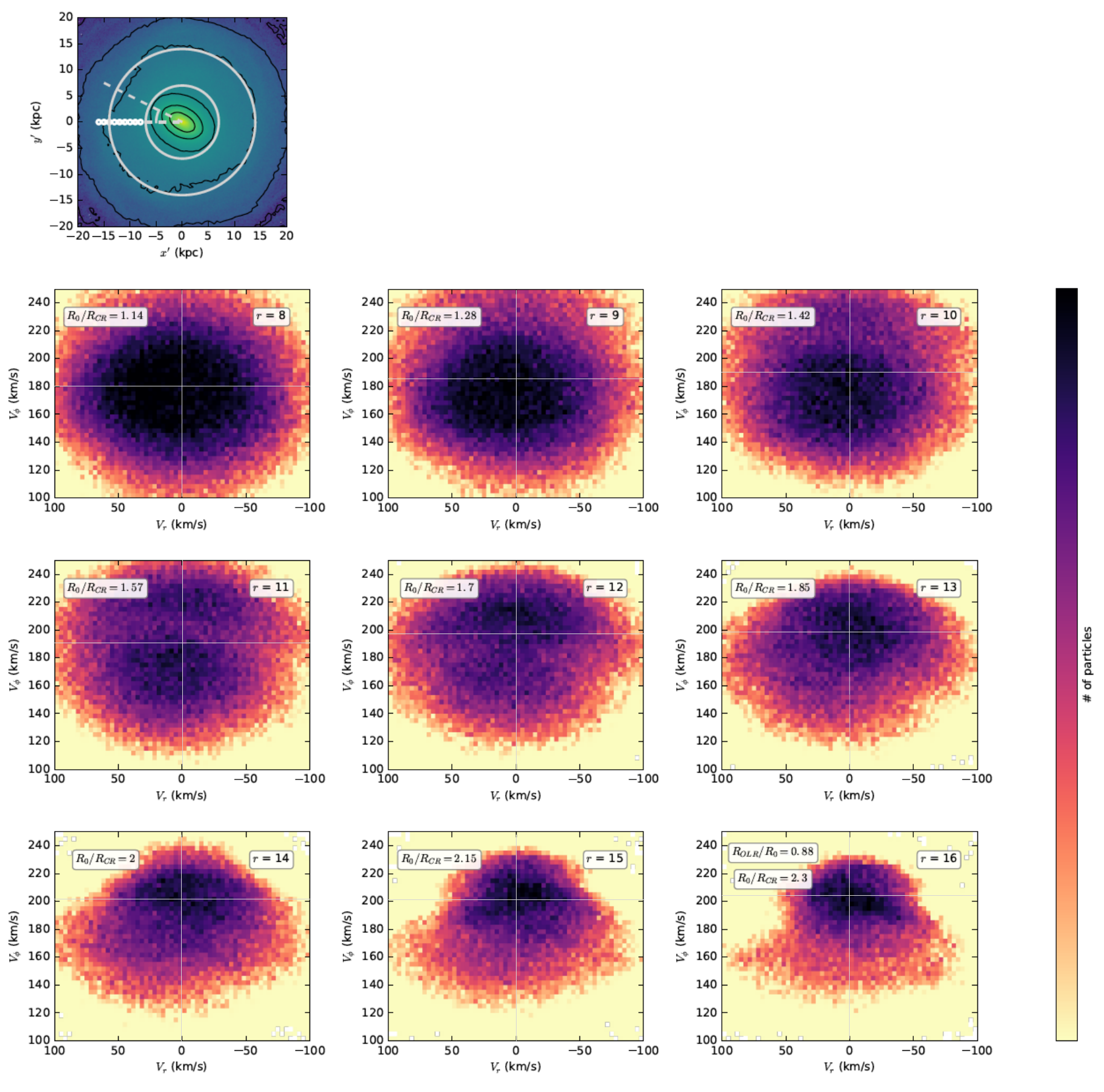}
\caption{$V_{\phi}$ vs $V_r$ for different ``Solar Neighbourhoods'' in the unscaled model. Top left panel shows the surface density of the unscaled model with the locations of the SN's indicated in small white cicles. The inner and outer grey circles show the corotation and OLR radius respectively. The bottom right panel corresponds to the fiducial OLR scenario.} 
\label{fig:vphivrallradii}
\end{figure*}

\begin{figure}
\centering
\includegraphics[width=0.99\linewidth]{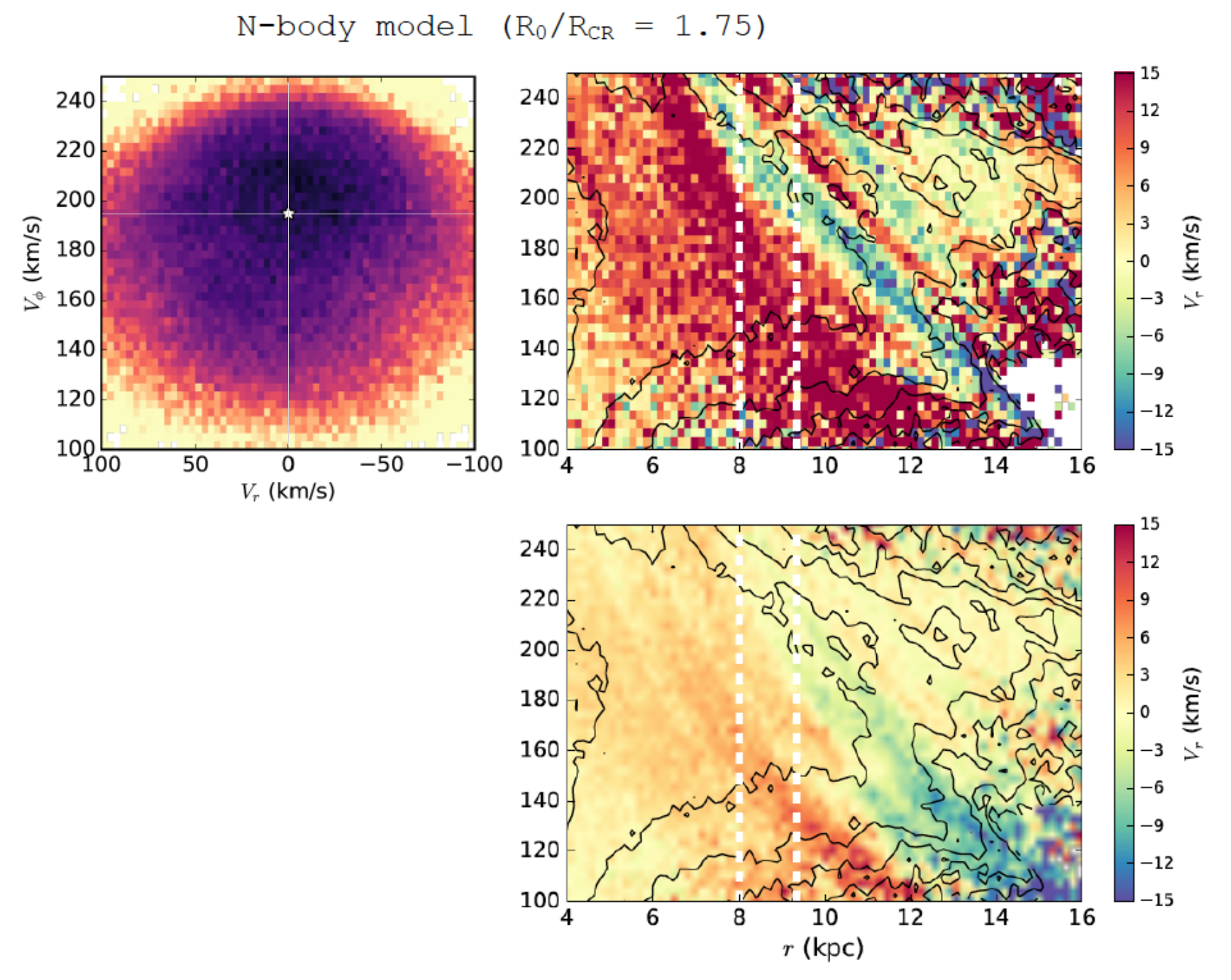}
\caption{$V_{\phi}$-$V_r$ and $V_{\phi}$-$r$ plane in a case where the SN is placed close to corotation (and $\Omega_{\rm p}$=39\,km/s/kpc). We see that in this case there is no strong asymmetry in $V_r$ in the $V_{\phi}$-$V_r$ plane caused by the bar.} 
\label{fig:corotationcase}
\end{figure}

There are two main competing scenarios for explaining the Hercules-like feature in the $V_{\phi} - V_r$ plane in terms of a bar-induced perturbation: the fast/short bar vs slow/long bar scenarios. In the former scenario the bar of the Milky Way is relatively short, of the order of 3.5-4\,kpc and has a pattern speed of $\sim$50-60km/s/kpc (as explored extensively both for the velocity distribution at the SN and gas dynamics inside the Solar radius e.g. \citealt{EnglmaierGerhard1999,Fux1999,Dehnen2000}). In this scenario the bimodality seen in the $V_{\phi} - V_r$ plane (i.e. the asymmetry in $V_r$ manifested as Hercules and `the horn') is due to the OLR of the bar, as in the scenario we have explored here. The other scenario postulates that, in accordance with recent measurements of the bar length in the literature (e.g. \citealt{WeggGerhard2013,Weggetal2015,Portailetal2017a}) and from studies of gas flows in the inner regions of the galaxy (e.g. \citealt{Sormanietal2015b,Lietal2016}), the bar of the Milky Way is longer, of the order of $\sim5$\,kpc. In this case the corotation must be further out and therefore the bar pattern speed will be lower. As a result, in this scenario the Hercules-like feature is due to orbits trapped at the corotation of the bar rather than OLR orbits (e.g. \citealt{PerezVillegasetal2017,Monarietal2018}). 

While recent studies of the gas flows in the Milky Way and measurement of the bar length are matched better by the slow/long bar model, the asymmetry in $V_r$ in the $V_{\phi} - V_r$ plane is better matched by the short/fast bar model: in the slow/long bar model the bimodality that gives rise to the Hercules-like feature is much weaker than in the short/fast bar case, and the asymmetry in $V_r$ is therefore also much weaker (see for example \citealt{PerezVillegasetal2017}). \cite{HuntBovy2018} attempted to alleviate this by introducing a stronger $m$=4 component in slow/long bar models in order to reproduce the strong bimodality in the $V_{\phi} - V_r$ plane. While introducing a stronger $m$=4 component, could indeed produce a stronger bimodality, there were other features which did not match up as well in their $V_{\phi} - V_r$ plane. It is still therefore under debate which model is a best fit to the overall properties of the Milky Way.

Here, we have explored the closed periodic orbits and trapped orbits close to the OLR in a model where the Solar vicinity is placed just outside the OLR, and we show that we can reproduce well the bimodality in the  $V_{\phi} - V_r$ plane, and provide an explanation for the strong `horn' feature, in terms of OLR resonant orbits. This strong asymmetry in $V_r$ in the $V_{\phi} - V_r$ plane translates to undulations in the $V_{\phi} - r$ plane which are also seen in the Gaia DR2 data.
 
While we do not explore in detail a model where the solar neighbourhood is close to corotation, we show what the  $V_{\phi} - V_r$ plane looks like at different radii in the disc, as well as close to the corotation of the bar, in Figure \ref{fig:vphivrallradii}. This Figure is constructed by stacking snapshots over a full rotation of the bar, in order to enhance features which are due to the bar and wash out features due e.g. spiral arms. We see that the only locations where a Hercules-like feature appears is close to the OLR (the model we explored above) and at about $R_{\rm 0}\sim1.7R_{\rm CR}$. In Figure \ref{fig:corotationcase} we show the $V_{\phi} - V_r$ plane at $R_{\rm 0}\sim1.75R_{\rm CR}$; if we rescaled this model to place the SN at 8\,kpc, it would correspond to a long/slow bar scenario with pattern speed 39\,km/s/kpc. We see that while a Hercules-like feature, i.e. a slight asymmetry at positive $V_r$ at low $V_{\phi}$'s is evident in the $V_{\phi} - V_r$ plane, it is not as strong as in the OLR case and there is no prominent `horn' feature. We note also that there is a relatively strong $m=4$ mode in our model, as shown in the right panel of Figure \ref{fig:rotcurve}, indicating that it is not the lack of the $m=4$ mode which makes the bimodality very weak in our simulation. This lack of a strong asymmetry in $V_r$, leads to a lack of `undulations' in the $V_{\phi} -r -V_r$ plane at radii corresponding to the SN (see the top and bottom panels on the right of Figure \ref{fig:corotationcase} for the $V_{\phi}$-$r$-$V_r$ space of a single snapshot and for stacked snapshots over one bar rotation period, respectively). We see therefore that the strong transition from outward to inward moving $V_r$ is a consequence of this strong bimodality in the $V_{\phi} - V_r$ plane and is a natural outcome of a case in which the SN is placed close to the OLR, while it is not reproduced in a case where the SN is placed closer to the corotation. 

If we therefore attempt to explain the strong asymmetry in $V_r$ in the $V_{\phi}$-$V_r$ plane (i.e. `the horn' vs Hercules) as being due to the bar, we must conclude that the most natural interpretation is one in which the bar is short and fast, since a slow/long bar does not reproduce the necessary features. However, the strong asymmetry in $V_r$, i.e. Hercules and `the horn', could be due to another mechanism, such as spiral arms(e.g. see \citealt{Huntetal2018}; \citealt{Sellwoodetal2018}; \citealt{Michtchenkoetal2018}; \citealt{Quillenetal2018}) and as such needs further investigation. In such a scenario, i.e. if we expect the bar of the Milky Way to be a slow/long bar, then we should see a Hercules and horn-like feature along with a long prominent ridge in the $V_{\phi}$-$r$ plane at larger radii, which would correspond to the OLR in such a scenario. Currently Gaia DR2 data does not seem to suggest such features, while the longest ridge in the $V_{\phi}-r$ plane appears to be the one associated to Hercules (see \citealt{Kawataetal2018}). However, the data is quite noisy at larger radii still and so future data releases might shed more light on this issue (see however \citealt{Monarietal2018} who claim to find a feature related to the bar OLR at larger $V_{\phi}$ when exploring Gaia DR2 in action space).

\section{Summary \& Conclusions}
\label{sec:summary}

The second Gaia data release revealed a plethora of substructures in the disc of the Milky Way -- such as ridges in the $V_{\phi}$-$r$ plane and undulations of $V_r$ in this plane -- as well as bringing into sharper focus the well-studied streams in the $V_{\phi}$-$V_r$ plane in the Solar vicinity. 
In order to study the origin and relation between these features, we make use of an N-body simulation -- which has been shown to be a good fit to the inner Milky Way and bulge (e.g. \citealt{Fragkoudietal2018}) -- and explore in detail the orbital structure in a scenario in which the Solar Neighbourhood is placed outside the OLR of the bar. To do this, we carry out a spectral analysis of the orbits and calculate the stable closed periodic orbits as well as trapped librating orbits around them;
we find that:

\begin{enumerate}
\item the most prominent and long-lived ridge formed in the $V_{\phi}$-$r$ plane is due to the OLR of the bar and is populated by elongated orbits, related to the $x_1(1)$ and $x_1(2)$ families
\item the particles trapped at the OLR resonance carry out small excursions in energy and angular momentum and are generally confined to a narrow region of the $E-L_z$ plane
\item the OLR ridge in the $V_{\phi}$-$r$ plane translates to streams in the $V_{\phi}$-$V_r$ plane at the Solar vicinity, namely the Hercules stream and the feature often referred to as `the horn'
\item `the horn', i.e. the feature which extends to negative $V_r$ in the $V_{\phi}$-$V_r$ plane is a result of orbits with small libration amplitudes around the $x_1(1)$ family (see also \citealt{Dehnen2000})
\item the Hercules stream on the other hand is made up of orbits which librate around both the $x_1(1)$ and $x_1(2)$ families with larger libration amplitudes
\item the presence of Hercules and `the horn' leads to a strong $V_r$ asymmetry in the $V_{\phi}$-$V_r$ plane. This asymmetry gives rise to one of the undulations seen in $V_r$ in the $V_{\phi}$-$r$ plane, and is a natural outcome of a model where the SN is placed just outside the OLR
\item the variation in the libration amplitude of orbits which populate different parts of the $V_{\phi}$-$V_r$ plane will have consequences on whether they originate predominantly in a hot or cold disc population. We show that for a given $V_{\phi}$, the ratio of particles from cold/hot populations changes as a function of $V_r$. Since in the MW disc, populations with different kinematics will also likely have different metallicities, this could have implications on the metallicity distribution in the $V_{\phi}$-$V_r$ plane.
\end{enumerate}

We also briefly explore the $V_{\phi}$-$r$ and $V_{\phi}$-$V_r$ planes closer to the bar corotation in the model, to compare and contrast with what we obtain at the OLR. We see that, as has been shown previously in the literature, a model where the SN is close to the corotation of the bar can produce a slightly skewed velocity distribution to positive $V_r$'s, i.e. to form a Hercules-like feature (e.g. \citealt{PerezVillegasetal2017}). This feature is however not as prominent as in the case where the SN is placed outside the OLR  (see also \citealt{HuntBovy2018}). In addition, in the case where the SN is placed close to the CR, there is no strong `horn'-like feature in the $V_{\phi}$-$V_r$ plane. This lack of a strong asymmetry in $V_r$ in the $V_{\phi}$-$V_r$ plane also leads to a lack of undulations in the $V_{\phi}-r - V_r$ space. 
We therefore see that Hercules and `the horn', i.e. the coexistence at the SN of an inward and outward moving feature in the $V_{\phi}$-$V_r$ plane, are naturally induced due to perturbations from the OLR of the bar, while a slow/long bar \emph{alone} cannot induce these features. It remains to be seen if the combination of a slow/long bar model together with other mechanisms, such as spirals or external perturbations, might be able to reproduce all features seen in the $V_{\phi}$-$V_r$-$r$ space in and around the Solar Neighbourhood. In such a case, we expect to find the OLR signatures, i.e. Hercules and `the horn', at larger galactocentric radii, while the ridge associated to them in the $V_{\phi}$-$r$ plane should be a long and prominent ridge. Further data releases from Gaia will be able to shed more light on this issue and resolve the degeneracy between the various scenarios for the bar pattern speed.

\section*{Acknowledgements}
FF thanks C. Damiani for helpful discussions and comments on earlier stages of this work.
SAK acknowledges support from the Russian Science Foundation, project no. 19-72-20089. 
This work has made use of data from the European Space Agency (ESA) mission
{\it Gaia} (\url{https://www.cosmos.esa.int/gaia}), processed by the {\it Gaia}
Data Processing and Analysis Consortium (DPAC,
\url{https://www.cosmos.esa.int/web/gaia/dpac/consortium}). Funding for the DPAC
has been provided by national institutions, in particular the institutions
participating in the {\it Gaia} Multilateral Agreement. This work was granted access to the HPC resources of CINES under the allocation A0020410154 made by GENCI and has been supported by the  ANR (Agence Nationale de la Recherche) through the MOD4Gaia project (ANR-15-CE31-0007, P.I.: P. Di Matteo). MCS acknowledges support from the Deutsche Forschungsgemeinschaft via the Collaborative Research Centre (SFB 881) ``The Milky Way System'' (sub-projects B1 and B8).

\bibliographystyle{mn2e}
\bibliography{References}

\appendix
\section{Details of the setup of the N-body model}
\label{sec:appendix}

The initial conditions of the model are obtained using the algorithm of \cite{Rodionovetal2009}, the so-called ``iterative'' method.
The algorithm constructs equilibrium phase models for stellar systems, thus avoiding the problem of the initial relaxation process often observed in N-body models of discs. This is achieved using a constrained evolution, so that the equilibrium solution has a number of desired parameters. In our case we impose the density distributions of the discs, which are described by a Miyamoto-Nagai profile \citep{BT2008}, where each disc has a characteristic radius $r_D$ given in Table \ref{tab:info}. The velocity dispersion is let to evolve unconstrained, with the requirement that the initial conditions (ICs) generated are in equilibrium. 

The time integration algorithm used is a recently developed parallel MPI Tree-code which takes into account the adaptive spatial decomposition of particle space between nodes. The multi-node Tree-code is based on the 256-bit AVX instructions which significantly speed up the floating point vector operations and sorting algorithms (Khoperskov et al. in prep). In total we employ 1.5$\times$$10^7$ particles in the model, $10^7$ in the disc and 5$\times$$10^6$ in the dark matter halo. The tolerance parameter of the tree-code is $\theta$ = 0.7, the time step is $\Delta t$ = 2 $\times$ 10$^5$ and for smoothing we use a Plummer potential with $\epsilon$ = 50\,pc.

\begin{figure*}
\centering
\includegraphics[width=0.99\textwidth]{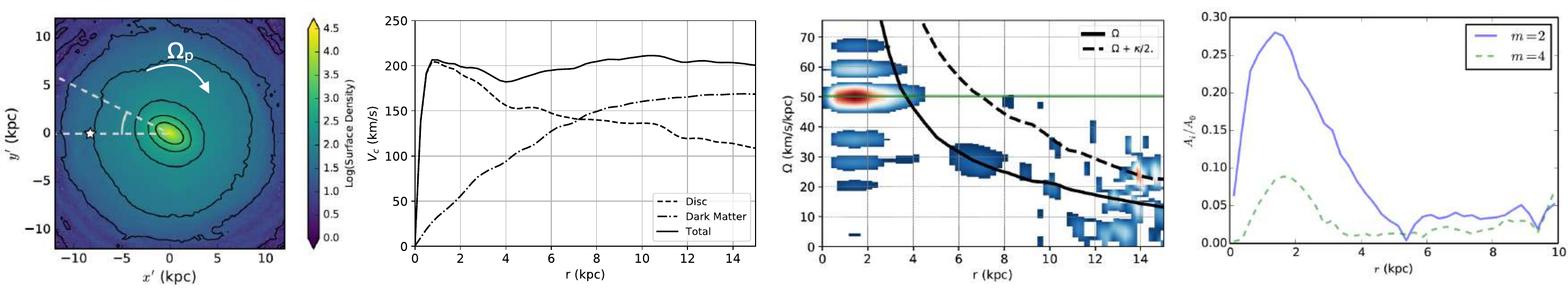}
\caption{\emph{Left:} Surface density of the model. The white star indicates the Sun's position and the dashed lines the angle between the galactocentric line of sight and the bar's semi-major axis, which in the fiducial model is 30 degrees. \emph{Middle Left:} Total circular velocity curve of the model (solid line), decomposed into the stellar disc (dashed line) and dark matter halo (dot-dashed line). \emph{Middle Right:} Spectrogram for the $m$=2 Fourier component of the surface density with $\Omega$($r$) (solid) and $\Omega$($r$)+$\kappa$/2 (dashed) overplotted. The horizontal solid green line indicates the bar pattern speed $\Omega_{\rm p}$. \emph{Right:} Strength of the $m=2$ and $m=4$ Fourier modes in the model.} 
\label{fig:rotcurve}
\end{figure*}

\label{lastpage}

\end{document}